\documentclass[11pt,a4paper]{article}
\pdfoutput=1
\usepackage[pdftex]{graphics}
\usepackage{jheppub}
\usepackage{amsmath,amssymb,amsfonts}
\usepackage{enumitem}
\usepackage{multirow}
\usepackage{array,booktabs}
\usepackage{tikz}
\usepackage{slashed}
\usepackage{verbatim}
\usepackage{float}
\usepackage{graphicx}
\usepackage{subcaption}
\usepackage{empheq}
\usepackage{cancel}
\usepackage{empheq}
\usepackage{comment}
\usepackage{enumitem}
\usepackage{titlesec}

\newlength{\oldfootskip}
\usepackage{tikz,subcaption}
\usetikzlibrary{hobby}
\usetikzlibrary{shapes.misc}
\tikzset{cross/.style={very thick, cross out, draw=black, fill=none, minimum size=3*(#1-\pgflinewidth), inner sep=0pt, outer sep=0pt}, cross/.default={3pt}}

\DeclareCaptionJustification{rjustified}{\rjustified}
\makeatletter
\newcommand\rjustified{%
  \let\\\@fillcr
  \leftskip\z@\@plus -1fil
  \rightskip\z@\@plus 1fil
  \parfillskip\z@\@plus 0fil\relax
}
\makeatother

\hypersetup{
  linktocpage,
  colorlinks  = true, %Colours links instead of ugly boxes
  urlcolor    = blue, %Colour for external hyperlinks
  linkcolor   = blue, %Colour of internal links
  citecolor   = blue %Colour of citations
}

\titleformat*{\section}{\Large\bfseries}
\titleformat*{\subsection}{\large\bfseries}

\def\o{\mathcal{O}}
\def\ap{\alpha'}
\def\Tr{\text{Tr}}
\def\R{\mathcal{R}}
\def\disc{\text{Disc}_s}
\def\stil{\tilde{s}}
\def\ttil{\tilde{t}}
\def\util{\tilde{u}}
\def\FS{\widehat{F}}

\newcommand\analytic[1]{#1}
\newcommand\leading[1]{#1}
\newcommand\sleading[1]{#1}

\usepackage{pdfpages}
\usepackage{pdflscape}

\usepackage{array}
\newcommand{\PreserveBackslash}[1]{\let\temp=\\#1\let\\=\temp}
\newcolumntype{C}[1]{>{\PreserveBackslash\centering}p{#1}}
\newcolumntype{R}[1]{>{\PreserveBackslash\raggedleft}p{#1}}
\newcolumntype{L}[1]{>{\PreserveBackslash\raggedright}p{#1}}

\title{Non-analytic terms of string amplitudes \\ from partial waves}
\author{Yu-tin Huang,$^{a}$}
\author{Hynek Paul,$^{b}$}
\author{Michele Santagata$^{a}$}

\affiliation{$^a$Department of Physics, National Taiwan University, Taipei 10617, Taiwan \\ $^{b}$Instituut voor Theoretische Fysica, KU Leuven, Celestijnenlaan 200D, B-3001 Leuven, Belgium}
\abstract{
We describe a general formalism based on the partial-wave decomposition to compute the iterative $s$-channel discontinuity of four-point amplitudes at any loop order. As an application, we focus on the low-energy expansions of type I and II superstring amplitudes. Besides providing new results for their leading and sub-leading logarithmic contributions beyond genus one, our approach elucidates the general structure of non-analytic threshold terms. In the case of open strings, the use of orthogonal colour projectors allows us to efficiently compute all contributions from different worldsheet topologies at a given loop order.}

\begin{document}
\maketitle
\newpage
%%%%%%%%%%%%%%%%%%%%%%%%%%%%%%%%%%%%%%%%%%%%%%%%%%%%%%%%%%%%%%%%%%%%%%%%%%%%%%%%%%%%%%%%%%
\section{Introduction}
%%%%%%%%%%%%%%%%%%%%%%%%%%%%%%%%%%%%%%%%%%%%%%%%%%%%%%%%%%%%%%%%%%%%%%%%%%%%%%%%%%%%%%%%%%
The four-point massless amplitudes for Type-I, Type-II and Heterotic strings were given at tree-level in the seminal papers that defined these theories. Achieving higher-loop results has been a lengthy but steady progress. Starting from the four-point tree-level genus one amplitude computed in~\cite{Green:1982sw}, the genus-two amplitude for four massless NS bosons was computed for Type II and Heterotic strings in the RNS formulation in~\cite{DHoker:2005vch, DHoker:2005dys}, whereas fermions were incorporated in the pure-spinor formulation in~\cite{Berkovits:2005df, Berkovits:2005ng}. At genus three, partial results include the verification of vanishing three-point amplitudes expected from non-renormalization theorems~\cite{Matone:2008td, Grushevsky:2009eqd}, and the leading low energy limit of the three-loop amplitude in type II theories~\cite{Gomez:2013sla}. Recently, a proposal for the full three-loop answer was given~\cite{Geyer:2021oox}, where one starts from the field theory limit and reverse-engineers the full fledged string theory result.
Beyond three loops only non-renormalization theorems for gravitons and gluons are known~\cite{Berkovits:2006vc, Berkovits:2009aw}.\footnote{See~\cite{Berkovits:2022ivl} for a recent review.}
Despite these efforts, a systematic understanding of the genus two and three low-energy expansion is still lacking.
Given the scarcity of explicit results, any partial multi-loop result will be a valuable reference for future computations.

In this paper we focus on the \textit{leading logarithms} of higher-loop (string) amplitudes. More precisely, at $L$-loops in string perturbation theory there will be a leading logarithmic contribution to the low-energy expansion of the amplitude $\mathcal{A}^{(L)}$
of the schematic form
\begin{align}\label{eq:max_log}
	\mathcal{A}^{(L)}\vert_{\ap^{4L+n}}\supset f^{(L,n)}(s,t)\log^L(-s)+\text{(crossing)}\,.
\end{align}
The aim of this work is to explain how to compute the leading-log coefficient functions $f^{(L,n)}(s,t)$ in an efficient and systematic manner, to any desired order in the loop- and $\ap$-expansion. Having explicit results for these coefficients provides a first step towards a potential partial resummation of string theory amplitudes. 

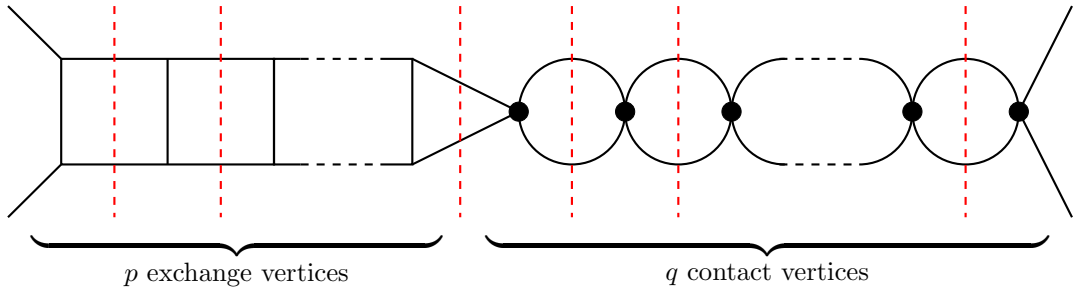
\begin{figure}
\begin{center}
\begin{tikzpicture}[scale=0.7]
\def\shift{0.1};
\def\shiftt{0.2};
\draw[thick] (-1,-1)--(0,0);
\draw[thick] (0,2)--(-1,3);
\draw[thick] (0,0)--(0,2)--(4,2)--(4,0)--(0,0);
\draw[thick] (2,0)--(2,2);
\draw[thick] (4.5,0)--(4,0)--(4,2)--(4.5,2);
\draw[thick,dashed] (4.5,0)--(6,0);
\draw[thick,dashed] (4.5,2)--(6,2);
\draw[thick] (6,0)--(6.5+\shift,0)--(6.5+\shift,2)--(6,2);
\draw[thick] (6.5+\shift,2)--(8.5+\shift,1)--(6.5+\shift,0);
\draw[thick] (9.5+\shift,1) circle (1);
\draw[thick] (11.5+\shift,1) circle (1);
\draw[thick] (13.5+\shift,0) arc (270:90:1);
\draw[thick,dashed] (13.5,0)--(15,0);
\draw[thick,dashed] (13.5,2)--(15,2);
\draw[thick] (15,0) arc (-90:90:1);
\draw[thick] (17,1) circle (1);
\draw[thick] (19,-1)--(18,1)--(19,3);
%higher-derivative vertices
\filldraw (8.5+\shift,1) circle (0.18);
\filldraw (10.5+\shift,1) circle (0.18);
\filldraw (12.5+\shift,1) circle (0.18);
\filldraw (16,1) circle (0.18);
\filldraw (18,1) circle (0.18);
%red cuts
\draw[thick,dashed,red] (1,3)--(1,-1);
\draw[thick,dashed,red] (3,3)--(3,-1);
\draw[thick,dashed,red] (7.5,3)--(7.5,-1);
\draw[thick,dashed,red] (9.5+\shift,3)--(9.5+\shift,-1);
\draw[thick,dashed,red] (11.5+\shift,3)--(11.5+\shift,-1);
\draw[thick,dashed,red] (17,3)--(17,-1);
\end{tikzpicture}\vspace{-0.9cm}
{\Large
\begin{align*}
	\hspace{0.00cm}\underbrace{\hspace{5.4cm}}_{ p\text{ exchange vertices}}\quad\underbrace{\hspace{7.4cm}}_{q\text{ contact vertices}}
\end{align*}}%
\end{center}\vspace{-0.4cm}
\caption{Iterative $s$-channel cut of a $L=p+q-1$ loop diagram with $p$ exchange and $q$ (higher-derivative) contact vertices. Only diagrams of the above type which allow $L$ number of two-particle cuts will contribute to the leading log, i.e. the $\log^L(-s)$ coefficient at $L$-loops.}\label{fig:s_cut}
\end{figure}

At genus one, different approaches have been explored to study these non-analytic contributions in the past: the traditional unitarity-cut based approach of e.g. \cite{Green:1999pv, Green:2008uj, Alday:2018pdi, DHoker:2019blr}, a method based on effective field theory for string interactions \cite{Edison:2021ebi}, and more recently computations directly on the string worldsheet \cite{Eberhardt:2022zay}, valid at finite $\ap$.
In order to generalize to higher genera, we utilize the unitarity-cut approach, where the leading-log coefficient functions are computed by $L$ iterative $s$-channel cuts, with tree-level amplitudes inserted on either side of each cut (see Figure \ref{fig:s_cut}).
The main improvement employed in this paper compared to the traditional method is to expand the tree-level amplitude into partial waves:
\begin{align}
	\mathcal{A}(s,\cos\theta) = \frac{1}{s^{\frac{d-4}{2}}}\sum_{\ell\text{ even}} a_\ell (s)P_{\ell}^{\frac{d-3}{2}}(z)\,, \quad z=\cos \theta=1+\frac{2 t}{s}\,.
\end{align}
Here $P_{\ell}^{\frac{d-3}{2}}(z)$ are the $d$-dimensional partial-wave polynomials (defined later in \eqref{eq:partial_wave_def}), which are a function of the scattering angle $\theta$ parametrised by $z$, and $a_\ell(s)$ denote the corresponding partial-wave coefficients. Note that while such an expansion is only possible for scalar amplitudes, for the string amplitudes considered in our work the nature of maximal supersymmetry (susy) allows us to factor out a universal prefactor that solves the susy Ward identities. The remaining functions of pure Mandelstam variables are then indeed susceptible to the expansion above. 

Importantly, due to the fact that the partial-wave polynomials are orthogonal with respect to the two-particle phase-space integrals $d\Omega^{(2)}_{\ell_1,\ell_2}$ (given in \eqref{eq:omega2}),
\begin{equation}\label{eq: Orthog0}
 \frac{1}{s^{\frac{d-4}{2}}} \int d\Omega_{\ell_1,\ell_2}^{(2)} \, P_\ell^{\frac{d-3}{2}} (z_{\text{f}})\,P_{\ell'}^{\frac{d-3}{2}} (z_{\text{i}}) =  P_\ell^{\frac{d-3}{2}}(z)\,\delta_{\ell,\ell'} \,,
\end{equation}
the iterative integral is translated into a simple multiplication of partial waves at fixed spin $\ell$.\footnote{ Indeed such an approach was already considered for the computation of leading infrared logarithms for ChPT in~\cite{Koschinski:2010mr}.} Here $z_{\text{i}}$ and $z_{\text{f}}$ parametrise the initial and final scattering angles, see the discussion around equation \eqref{eq:Orthogpart} for a more detailed explanation. In particular, since beyond leading order in the $\alpha'$-expansion the tree-amplitude contributes to a finite number of partial waves only, equation (\ref{eq: Orthog0}) immediately implies that the leading-log coefficient at any loop-order will also be computed by finitely many partial waves. Lastly, it is worth pointing out that this method is straightforwardly applicable to any spacetime dimension $d$. For instance, the partial-wave decomposition of the open string amplitude in $d=8$ was recently applied in the context of the flat-space limit of one-loop AdS amplitudes \cite{Paul:2023zyr}.

%%%%%%%%%%%%%%%%%%%%%%%%%%%%%%%%%%%%%%%%%%%%%%%%%%%%%%%%%%%%%%%%%%%%%%%%%%%%%%%%%%%%%%%%%
%%%%%%%%%%%%%%%%%%%%%%%%%%%%%%%%%%%%%%%%%%%%%%%%%%%%%%%%%%%%%%%%%%%%%%%%%%%%%%%%%%%%%%%%%

%%%%%%%%%%%%%%%%%%%%%%%%%%%%%%%%%%%%%%%%%%%%%%%%%%%%%%%%%%%%%%%%%%%%%%%%%%%%%%%%%%%%%%%%%
\subsubsection*{Overview and summary of main results}
For readers with little time, we now give a brief overview of our main results.

In \textbf{Section \ref{sec:partial_waves}}, we begin with a general discussion of how the leading logarithmic discontinuity of an $L$-loop amplitude can be computed from iterated $s$-channel unitarity cuts. The leading log in fact only depends on the \textit{tree-level} vertices entering the cut. This is illustrated pictorially in Figure \ref{fig:s_cut}. As mentioned before, employing the partial-wave decomposition greatly simplifies this computation. The leading-log coefficient $f^{(L)}(s,t)$ is then simply given by
\begin{equation}
f^{(L)}(s,t) = \Big(-\frac{1}{4\pi}\Big)^L \frac{1}{s^{\frac{d-4}{2}}} \sum_{\ell\text{ even}} \big(a^{(0)}_{\ell}\big)^{L+1} P_{\ell}^{\frac{d-3}{2}}(z) \,,
\end{equation}
with $a^{(0)}_\ell$ being the tree-level partial-wave coefficients.

We then apply this formalism to two concrete examples: in \textbf{Section \ref{sec:closed_strings}}, we consider the low-energy expansion of the four-point graviton amplitude in type II string theory and derive new results for higher-genus non-analytic terms at small $\ap$. Explicit results up to genus 3 are given in Section \ref{sec:leading_logs_closed}. Moreover, the leading log partial-wave coefficients of the leading Regge trajectory are so simple that it is possible to perform an all-loop resummation. For odd powers $(\ap)^n$ for instance, we find
\begin{equation}
\sum_{L=0}^\infty \big[f^{(L,n)}(\stil,\ttil)\vert_{\tilde{t}^{n-3}}\big]\cdot\log^L(-\tilde{s}) = \frac{1}{8^{n-3}}\,\frac{2\,\zeta_{n}}{\Big(1 + \frac{8 \pi^2 \tilde{s}^4\log(-\tilde{s})}{(n-2)_6}\Big)^2}\,, \quad n\geq3\text{ and odd}\,,
\end{equation}
where $(x)_n \equiv \Gamma(x+n)/\Gamma(x)$ denotes the Pochhammer symbol. This corresponds to the all-loop renormalisation of the leading Regge trajectory due to supergravity effects.

In \textbf{Section \ref{sec:general_log_structure}}, we explain how our method also elucidates the general structure of logarithmic terms in the low-energy expansion. Taking the known higher-genus analytic terms in type IIB string theory as input, we compute the coefficients of certain sub-leading log's at genus 2 and 3. For instance, the first few terms in the low-energy expansion of the genus-two amplitude take the form
\begin{align}
\begin{split}
	A^{(2)} &= \ap^3\bigg(\frac{\pi^4}{1080}\sigma_2 + \frac{\pi^4}{1440}\,\sigma_3 \\ &\quad\qquad-\bigg[\frac{2\pi^4}{135}\Big(\stil^4\log(-\stil)+\ttil^4\log(-\ttil)+\util^4\log(-\util)\Big)+(\texttt{analytic})\bigg] \\
	&\quad\qquad+A^{(2,0)}(\stil,\ttil)+\ldots\bigg),
\end{split}
\end{align}
where the middle line is a new prediction which in fact precedes the two-loop supergravity amplitude $A^{(2,0)}(\stil,\ttil)$ by one order in $\ap$. For higher-order terms in $\ap$, see our full result \eqref{eq:genus_2_final}, and an analogous expansion of the genus-three amplitude is given in \eqref{eq:genus_3_final}.
These considerations also lead us to conjecture that the (currently unknown) genus-2 correction of the unprotected $\partial^{12}\R^4$ higher-derivative term is proportional to $\zeta_3$, see \textbf{Conjecture~\ref{conj}}.

In \textbf{Section \ref{sec:open_strings}}, as a second application of our method, we consider the type I open string amplitude. The new feature in this case is the non-trivial colour structure, which is dealt with by introducing projectors onto irreducible representations of the gauge group. In particular, this has the advantage that contributions associated to different worldsheet topologies are computed all at once. Our main results for open string amplitudes are the leading log predictions computed in Section \ref{sec:leading_logs_open}.

Lastly, in \textbf{Section \ref{sec:conclusions}}, we discuss our results and point out possible future directions.

%%%%%%%%%%%%%%%%%%%%%%%%%%%%%%%%%%%%%%%%%%%%%%%%%%%%%%%%%%%%%%%%%%%%%%%%%%%%%%%%%%%%%%%%%%
\section{Partial-wave expansions for unitarity cuts}\label{sec:partial_waves}
%%%%%%%%%%%%%%%%%%%%%%%%%%%%%%%%%%%%%%%%%%%%%%%%%%%%%%%%%%%%%%%%%%%%%%%%%%%%%%%%%%%%%%%%%%
We will be interested in the coefficient of  $\log^L(-s)$ at $L$-loops, which is accessible via $L$-iterated massless $s$-channel cuts. Here we keep the discussion general and keep the number of dimensions $d$ arbitrary. All we need to assume is that the loop expansion is controlled by a dimensionless coupling. Beginning at one loop, we have
\begin{equation}\label{disc_one-loop}
\text{Disc}_s [\mathcal{A}^{(1)}(p_1,p_2,p_3,p_4)]= - \frac{1}{4\pi} \int d\Omega_{\ell_1,\ell_2}^{(2)}\,\, \mathcal{A}^{(0)}(p_1, p_2, \ell_1,  \ell_2)\mathcal{A}^{(0)}({-}\ell_1, {-}\ell_2, p_3, p_4)\,,
\end{equation}
where $d\Omega_{\ell_1,\ell_2}^{(2)}$ is the $2$-particle phase space integral,
\begin{equation}\label{eq:omega2}
\int d\Omega_{\ell_1,\ell_2}^{(2)} = (2\pi)^d \int d^d\ell_1d^d\ell_2 \,\delta^{(+)}(\ell_1^2) \delta^{(+)}(\ell_2^2)\delta^d(\ell_1{+}\ell_2{-}p_1{-}p_2)\,,
\end{equation}
with $\delta^{(+)}(\ell^2) =\delta (\ell^2) \theta (\ell_0)$, and the $s$-channel discontinuity is defined as the difference between the amplitude above and below the massless $s$-channel cut (starting at $s=0$) in the complex $s$-plane:
\begin{equation}\label{defddisc}
\text{Disc}_s[\mathcal{A}(s,t)]\equiv\frac{1}{2\pi i}\big(\mathcal{A}(s-i0,t)-\mathcal{A}(s+i0,t)\big)\,.
\end{equation}
Note that our definition of discontinuity is normalised such that $\disc\log(-s)=1$. The presence of this massless cut tells us that the one-loop amplitude can be written as $\mathcal{A}^{(1)} = f(s,t) \log(-s)+g(s,t)$, where the unitarity equation \eqref{disc_one-loop} above determines the function $f(s,t)$. Note, however, that the functions $f$ and $g$ can be in itself non-analytic, i.e. the only distinguishing feature is that $g(s,t)$ does not have a branch-cut in the $s$-channel. In fact, if the tree-level amplitude contains a massless pole, the phase-space integral can generate further $\log(t/s)$ terms (and possibly other transcendental functions) in $f(s,t)$.

Analogously, at two loops we have
\begin{align}
\begin{split}
\text{Disc}_s[\mathcal{A}^{(2)}(p_1,p_2,p_3,p_4)] &= 2\cdot\Big(-\frac{1}{4\pi}\Big) \int d\Omega_{\ell_1,\ell_2}^{(2)}\, \mathcal{A}^{(1)}(p_1, p_2, \ell_1,  \ell_2)\mathcal{A}^{(0)}({-}\ell_1, {-}\ell_2, p_3, p_4)\\
&{+}\,\mathcal{N}_3 \int d\Omega_{\ell_1,\ell_2,\ell_3}^{(3)}\,\mathcal{A}^{(0)}(p_1, p_2, \ell_1,  \ell_2, \ell_3)\mathcal{A}^{(0)}({-}\ell_1, {-}\ell_2, {-}\ell_3,  p_3, p_4)\,,
\end{split}
\end{align}
where the 2 in the first line is a combinatorial factor, and $\mathcal{N}_3$ is some normalisation of the three-particle cut whose value is irrelevant for the purpose of this work.
In fact, the three-particle unitarity cut comprises of two tree amplitudes, thus taking another discontinuity on both sides only the first term survives, and one has:
\begin{eqnarray}
\begin{split}
\text{Disc}_s\text{Disc}_s[\mathcal{A}^{(2)}(p_1,p_2,p_3,p_4)]=2\cdot\Big(-\frac{1}{4\pi}\Big)^2  & \int d\Omega_{\ell_1,\ell_2}^{(2)}\int d\Omega_{\ell_3,\ell_4}^{(2)}\,\mathcal{A}^{(0)}(p_1, p_2, \ell_1,  \ell_2)\\
&\times\mathcal{A}^{(0)}({-}\ell_1, {-}\ell_2, \ell_3, \ell_4)\mathcal{A}^{(0)}({-}\ell_3, {-}\ell_4,  p_3, p_4)\,.
\end{split}
\end{eqnarray}
We hence see that the maximal massless $s$-channel discontinuity at $L$-loops is obtained via $L$ iterative cuts, and only depends on the four-particle tree amplitude. Thus equipped with the four-particle tree-level amplitude one can obtain the coefficient of $\log^L(-s)$ for arbitrary loop order:
\begin{eqnarray}\label{ddiscLloops}
\begin{split}
\text{Disc}_s^L [\mathcal{A}^{(L)}&(p_1,p_2,p_3,p_4)]= L!\cdot\Big(-\frac{1}{4\pi}\Big)^L \bigg( \prod_{i=0}^{L-1} \int d\Omega_{\ell_{2i+1}\ell_{2i+2}}^{(2)} \bigg)\\
&\mathcal{A}^{(0)}(p_1, p_2, \ell_1,  \ell_2)\mathcal{A}^{(0)}({-}\ell_1, {-}\ell_2, \ell_3, \ell_4)\cdots \mathcal{A}^{(0)}({-}\ell_{2L{-}1}, {-}\ell_{2L},  p_3, p_4)\,.
\end{split}
\end{eqnarray}

The two-particle cut in equation \eqref{ddiscLloops} can be computed efficiently by utilizing the fact that Gegenbauer polynomials are orthogonal with respect to the phase space measure. Let us expand the amplitude as 
\begin{align}\label{eq:partialwaveexp_general}
	\mathcal{A}(s,\cos\theta) = \frac{1}{s^{\frac{d-4}{2}}}\sum_{\ell\text{ even}} a_\ell (\tilde{s})P_{\ell}^{\frac{d-3}{2}}(z)\,, \quad z=\cos \theta=1+\frac{2 t}{s}\,,
\end{align}
where the factor of $s^{\frac{d-4}{2}}$ in the denominator ensures that the partial-wave coefficients $a_\ell$ are function of the dimensionless variable $\tilde{s}=\alpha' s$, where $\alpha'$ is some dimensionful scale of the EFT. In string theory, it corresponds to the string length. The polynomials $P_{\ell}^{\frac{d-3}{2}}(z)$ are given by
\begin{align}\label{eq:partial_wave_def}
	P_{\ell}^{\frac{d-3}{2}}(z) = 2^{2d-5}\pi^{\frac{d-3}{2}}(d+2\ell-3)\Gamma \big[\tfrac{d-3}{2} \big]C_{\ell}^{\frac{d-3}{2}}(z)\,,
\end{align}
where the partial waves $P_{\ell}^{\alpha}(z) $ are orthonormal polynomials with respect to the phase space measure (see e.g. \cite{Soldate:1986mk}):
\begin{equation}\label{eq:Orthogpart}
 \frac{1}{s^{\frac{d-4}{2}}} \int d\Omega_{\ell_1,\ell_2}^{(2)} \, P_\ell^{\frac{d-3}{2}} (z_{\text{f}})\,P_{\ell'}^{\frac{d-3}{2}} (z_{\text{i}}) =  P_\ell^{\frac{d-3}{2}}(z)\,\delta_{\ell,\ell'} \,,
\end{equation}
where $z_{\text{i}}=\cos\theta_{\text{i}}$ and $z_{\text{f}}=\cos\theta_{\text{f}}$ parametrise the initial and final scattering angles $\theta_{\text{i}}$ and $\theta_{\text{f}}$, respectively.\footnote{\label{foot}In the center of mass frame of the $p_1+p_2\rightarrow p_3+p_4$ scattering, $\theta_{\text{i}}$ is the angle between $\vec{p}_1$ and $\vec{\ell}_1$, while $\theta_{\text{f}}$ is the angle between $\vec{\ell}_1$ and $\vec{p}_3$. The total scattering angle $\theta$, related to $z$ as given in \eqref{eq:partialwaveexp_general}, is as usual the angle between $\vec{p}_1$ and $\vec{p}_3$.} By using the partial-wave decomposition \eqref{eq:partialwaveexp_general} for the tree-level amplitudes $\mathcal{A}^{(0)}$ and applying the above relation iteratively to the RHS of \eqref{ddiscLloops}, we obtain
\begin{align}\label{eq:disc_partial_waves}
\text{Disc}_s^L [\mathcal{A}^{(L)}(p_1,p_2,p_3,p_4)] = \frac{L!}{s^{\frac{d-4}{2}}} \Big(-\frac{1}{4\pi}\Big)^L \sum_{\ell\text{ even}} \big(a^{(0)}_{\ell}(\tilde{s})\big)^{L+1} P_{\ell}^{\frac{d-3}{2}}(z)\,,
\end{align}
where by $a_\ell^{(0)}$ we denote the partial-wave coefficients of the tree-level amplitude $\mathcal{A}^{(0)}$. Let us also expand the LHS of (\ref{ddiscLloops}) on $P_{\ell}^{\frac{d-3}{2}}(z)$ with the normalization in eq. (\ref{eq:partialwaveexp_general}),
\begin{equation}\label{ddiscLloops1}
\text{Disc}_s^L [\mathcal{A}^{(L)}(p_1,p_2,p_3,p_4)]=\frac{1}{s^{\frac{d-4}{2}}}\sum_{\ell\text{ even}} a^{\text{Disc},L}_\ell(\tilde{s}) P_{\ell}^{\frac{d-3}{2}}(z)\,.
\end{equation}
Matching both sides we obtain
\begin{equation}
\label{ddiscLloops2}
a_\ell^{\text{Disc},L}(\tilde{s}) =  L!\cdot\Big(-\frac{1}{4\pi}\Big)^L\big(a^{(0)}_{\ell}(\tilde{s})\big)^{L+1}  \,,
\end{equation}
giving us a simple relation between the partial-wave coefficients of the $L$-th discontinuity at $L$-loop order in terms of the tree-level coefficients $a_\ell^{(0)}$. Plugging this into \eqref{ddiscLloops1}, one can straightforwardly reconstruct the function in front of $\log^L(-s)$ via\footnote{
The factor of $L!$ in \eqref{ddiscLloops2} cancels against a $1/L!$ in \eqref{ddiscLloops1} coming from the iterated application of \eqref{defddisc} which yields $\text{Disc}_s^L [\log^L(-s)]=L!$.}
\begin{empheq}[box=\fbox]{equation}
\begin{aligned}\label{eq:logL_coeff}
f^{(L)}(s,t) = \Big(-\frac{1}{4\pi}\Big)^L \frac{1}{s^{\frac{d-4}{2}}} \sum_{\ell\text{ even}} \big(a^{(0)}_{\ell}(\tilde{s})\big)^{L+1} P_{\ell}^{\frac{d-3}{2}}(z)  \,.
\end{aligned}
\end{empheq}
As a consequence of the parity symmetry of Gegenbauer polynomials (they are invariant under $z\mapsto-z$ for even spin, corresponding to a $t\leftrightarrow u$ exchange), the functions $f^{(L)}(s,t)$ respect the $s$-channel crossing symmetry
\begin{align}\label{eq:crossing_tu}
	f^{(L)}(s,t) = f^{(L)}(s,u)\,.
\end{align}

Let us remark that the above construction is very general and applicable to any EFT. As suggested by Figure \ref{fig:s_cut}, the EFT can allow for both exchange and (higher derivative) contact vertices, as long as they are separated by a two-particle cut. In the following, we will apply our method to two concrete examples and compute new non-analytic terms in four-point string amplitudes in 10 dimensions, starting with the closed string case below. 

%%%%%%%%%%%%%%%%%%%%%%%%%%%%%%%%%%%%%%%%%%%%%%%%%%%%%%%%%%%%%%%%%%%%%%%%%%%%%%%%%%%%%%%%%%
\section{Application I: the type II closed string amplitude}\label{sec:closed_strings}
%%%%%%%%%%%%%%%%%%%%%%%%%%%%%%%%%%%%%%%%%%%%%%%%%%%%%%%%%%%%%%%%%%%%%%%%%%%%%%%%%%%%%%%%%%
As a first application of the partial-wave construction described above, let us consider the massless four-point amplitude in type II string theory. Each external particle is labelled by a momentum $k_i$ (with $k_i^2=0$) and a `super-polarisation vector' $\xi_i$ encoding the 256 degrees of freedom of the maximal supergravity multiplet. In string perturbation theory, the four-graviton amplitude admits a genus expansion of the form
\begin{align}\label{eq:genus_exp}
	\mathcal{A}(k_i,\xi_i) = \kappa_{10}^2\, \R^4 \, \sum_{h=0}g_s^{2h+2} A^{(h)}(s, t;\ap)\,,
\end{align}
where $\kappa_{10}^2\equiv2^6\pi^7\ap^4$ and $\mathcal{R}^4\equiv\mathcal{R}^4_{\xi_1\xi_2\xi_3\xi_4}(k_1,k_2,k_3,k_4)$ is an overall universal kinematical prefactor which carries all the dependence on the polarisation vectors $\xi_i$. Note that one can factor out an overall $\alpha'^3$ in $A^{(h)}$ to saturate the dimension of the amplitude,
\begin{align}
	A^{(h)}(s,t;\ap) = \ap^3\,A^{(h)}(\stil,\ttil)\,,
\end{align}
leaving behind a dimensionless function of $\tilde{s}\equiv\alpha' s$ and $\tilde{t}\equiv\alpha' t$. Lastly, an important property of the polarisation tensor is the `sewing relation' derived in \cite{Bern:1998ug}: due to maximal supersymmetry, the product of two $\mathcal{R}^4$ tensors in the unitarity cut has the self-replicating property 
\begin{align}\label{eq:sewing_relation}
	\sum_{\rm 2\, pt \,states}\mathcal{R}^4_{1,2\rightarrow {\rm 2\, pt }}\,\mathcal{R}^4_{ {\rm 2\, pt } \rightarrow 3,4} = s^4\,\mathcal{R}^4_{1,2\rightarrow 3,4}\,,
\end{align} 
where the sum is over all states in the supergravity multiplet.

%%%%%%%%%%%%%%%%%%%%%%%%%%%%%%%%%%%%%%%%%%%%%%%%%%%%%%%%%%%%%%%%%%%%%%%%%%%%%%%%%%%%%%%%%%
\subsection{From tree-level partial waves to leading log's at any genus}
As explained in Section \ref{sec:partial_waves}, all we need are the tree-level partial wave coefficients. To this end, consider the genus-zero contribution given by the Virasoro-Shapiro amplitude
\begin{align}\label{EFTExp}
	A^{(0)}(s,t;\ap) = \frac{64}{stu}\frac{\Gamma(1-\tilde{s}/4)\Gamma(1-\tilde{t}/4)\Gamma(1-\tilde{u}/4)}{\Gamma(1+\tilde{s}/4)\Gamma(1+\tilde{t}/4)\Gamma(1+\tilde{u}/4)} = \frac{64}{stu}\exp\bigg(\sum_{m=1}^\infty\frac{2\zeta_{2m+1}}{2m+1}\sigma_{2m+1}\bigg),
\end{align}
where we used $\sigma_n\equiv\tilde{s}^n+\tilde{t}^n+\tilde{u}^n$. Instead of directly expanding the whole amplitude on the partial-wave basis, we will first perform an $\alpha'$-expansion,
\begin{equation}\label{eq:A0_exp}
A^{(0)}(s,t;\ap)=\alpha'^3 \bigg(\frac{64}{ \tilde{s}\tilde{t}\tilde{u}}+ \sum_{n=3}^\infty  m_n(\tilde{s},\tilde{t}) \bigg),
\end{equation}
where we factored out an overall $\ap^3$ such that the expression in brackets is manifestly dimensionless. The $m_{n}(\tilde{s}, \tilde{t})$ are then homogeneous polynomials of degree $n-3$ in $\tilde{s}, \tilde{t}$, which are easily obtained by expanding the exponentiated rewriting of $A^{(0)}$ given in \eqref{EFTExp}. For concreteness, the first few terms beyond the supergravity contribution read
\begin{align}\label{eq:low_n_examples}
	m_3(\stil,\ttil)=2\zeta_3\,,\quad m_5(\stil,\ttil)=\frac{1}{16}\sigma_2\zeta_5\,,\quad m_6(\stil,\ttil)=\frac{1}{96}\sigma_3\zeta_3^2\,,\quad m_7=\frac{1}{512}\sigma_2^2\zeta_7\,,
\end{align}
corresponding to the tree-level $\R^4$, $\partial^4\R^4$, $\partial^6\R^4$ and $\partial^8\R^4$ higher-derivative corrections, respectively.

The advantage of performing the low-energy expansion is that, apart from the supergravity term, each higher-derivative term will only contribute to a finite number of partial waves due to the fact that the $m_{n\geq3}(\stil,\ttil)$ are purely polynomial.
Expanding on $P_{\ell}(z)$ as in \eqref{eq:partialwaveexp_general}, we define:\footnote{From now on we will only consider amplitudes in $d=10$ dimensions, and therefore drop the superscript $\frac{d-3}{2}=\frac{7}{2}$ in the partial waves: $P_{\ell}(z) \equiv P_{\ell}^{\frac{7}{2}}(z)$.}
\begin{align}\label{eq:tree-level_exp}
\ap^3\,\frac{64}{\tilde{s}\tilde{t}\tilde{u}}= \frac{1}{s^3}\sum_{\substack{\ell=0\\\ell\text{ even}}}^\infty\, \epsilon_\ell^{(0)}(\tilde{s})\,P_{\ell}(z)\,,\quad	\alpha'^3\, m_n(\tilde{s}, \tilde{t}) =\frac{1}{s^3}\sum_{\substack{\ell=0\\\ell\text{ even}}}^{2\lfloor \frac{n-3}{2}\rfloor}\, \epsilon_\ell^{(n)}(\tilde{s})\,P_{\ell}(z)\,,~~(n\geq3)
\end{align}
with the partial-wave coefficients $\epsilon_\ell^{(n)}$ being functions of dimensionless $\tilde{s}$. Here we use the notation $\epsilon_\ell$ instead of $a_\ell$ for the partial-wave coefficients in order to distinguish the $\kappa_{10}^2 \mathcal{R}^4 g_s^2$ subtracted partial-wave decomposition of $A^{(0)}$ from the decomposition of the full amplitude $\mathcal{A}$.\footnote{Of course, the 2-particle unitarity relation instructs us to take the full amplitude into consideration, so in the end we need to reinstate the correct factors of $\kappa_{10}^2 \mathcal{R}^4 g_s^{2h+2}$. We will comment on this later on again.}
The supergravity contribution $\epsilon_\ell^{(0)}$ has support on all even spins and takes the well-known form \cite{Caron-Huot:2018kta}
\begin{equation}\label{sugrashifts}
\text{SUGRA:}\qquad\epsilon_\ell^{(0)}= \frac{1}{2\pi^4(\ell+1)_6}\,.\hspace{3.5cm}
\end{equation}
On the other hand, the $\epsilon_\ell^{(n\geq3)}$ have only finite spin support: they are non-vanishing for even spins with $\ell\leq 2\lfloor \frac{n-3}{2}\rfloor$. For example, from the first few orders given in \eqref{eq:low_n_examples} we obtain
\begin{align}\label{treelshifts}
\begin{split}
\R^4:\qquad\epsilon_\ell^{(3)}&=\frac{\tilde{s}^3\zeta_3}{2^{11}\cdot105\pi^4}\,\delta_{\ell,0}\,,\\
\partial^4\R^4:\qquad\epsilon_\ell^{(5)}&=\frac{\tilde{s}^5\zeta_5}{2^{15}\cdot135\pi^4}\,\Big(\delta_{\ell,0}+\frac{1}{154}\delta_{\ell,2}\Big),\\
\partial^6\R^4:\qquad\epsilon_\ell^{(6)}&=\frac{\tilde{s}^6\zeta_3^2}{2^{16}\cdot945 \pi^4}\,\Big(\delta_{\ell,0}-\frac{1}{44}\delta_{\ell,2}\Big)\,,\\
\partial^8\R^4:\qquad\epsilon_\ell^{(7)}&=\frac{\tilde{s}^7\zeta_7}{2^{17}\cdot693 \pi^4}\,\Big(\delta_{\ell,0}+\frac{7}{520}\delta_{\ell,2}+\frac{1}{7800}\delta_{\ell,4}\Big)\,.\hspace{-1.5cm}
\end{split}
\end{align}
More generally, the tree-level partial-wave coefficients $\epsilon_\ell^{(n\geq3)}$ can be efficiently extracted for any order in $\ap$ upon noting that the polynomials $m_n(\stil,\ttil)$ can be written as
\begin{equation}\label{eq:m_decomp}
m_n(\tilde{s}, \tilde{t})  = \sum_{k=0}^{n-3} c_{n,k} \cdot \tilde{s}^{n-k-3}\, \tilde{t}^{k},
\end{equation}
for some constant coefficients $c_{n,k}$, with the partial-wave decomposition of a generic monomial $\stil^a\ttil^b$ being given by
\begin{align}\label{eq:pw_monomial}
& \ap^3\,\tilde{s}^a\,\tilde{t}^b = \frac{1}{s^3} \, \frac{\tilde{s}^{a+b+3}}{512\pi^4}\sum_{\ell=0}^\infty   \frac{(-1)^{b+\ell}\,\Gamma(b+1)\Gamma(b+4)}{\Gamma(b-\ell+1)\Gamma(b+\ell+8)} \, P_\ell (z) \,.
\end{align}
Note that the finite support in spin is entirely dictated by the exponent of $\ttil$: the monomial $\tilde{s}^a\,\tilde{t}^b$ contributes to (both even and odd) spins $\ell=0,1,\ldots,b$, with the truncation naturally provided by the denominator factor $\Gamma(b-\ell+1)$. Moreover, all odd spin contributions cancel in the sum \eqref{eq:m_decomp} due to the $m_n(\stil,\ttil)$ being symmetric under $\ttil\leftrightarrow\util$ exchange.\\

With the tree-level partial-wave decomposition in place, we are ready to apply the formalism described in Section \ref{sec:partial_waves}. To this end, let us denote the coefficient of $\log^h(-\stil)$ of the genus-$h$ amplitude $A^{(h)}$ by $f^{(h)}(s,t)$, i.e. we define
\begin{align}
	f^{(h)}(s,t;\ap)\equiv A^{(h)}(s,t;\ap)\vert_{\log^h(-\stil)}\,.
\end{align}
As for the amplitude itself, we can pull out an overall factor of $\ap^3$ to obtain a dimensionless function of $\stil$ and $\ttil$. The low-energy expansion of $f^{(h)}$ then takes the form
\begin{equation}\label{eq:fh_exp}
	f^{(h)}(s,t;\ap)=\ap^3 f^{(h)} (\tilde{s},\tilde{t})= \ap^3\sum_{n=0}^\infty f^{(h,n)}(\tilde{s},\tilde{t}) = \sum_{n=0}^\infty (\ap)^{4h+n} f^{(h,n)}(s,t) \,,
\end{equation}
with the $f^{(h,n)}(\tilde{s},\tilde{t})$ being homogeneous polynomials of order $4h+n-3$.\footnote{As we will see momentarily, this is actually only true for $n>0$. For $n=0$, $f^{(h,0)}(\tilde{s},\tilde{t})$ is a rather non-trivial transcendental function which arises from a resummation of infinitely many partial waves. This is a direct consequence of the infinite spin support of the supergravity partial-wave coefficients \eqref{sugrashifts}.}
In words, the above notation for $f^{(h,n)}$ is such that the superscript $(h,n)$ labels the genus $h$ and the order in $\ap$ is given by $4h+n$.

We then consider the partial-wave expansion order by order in $\ap$:
\begin{equation}\label{eq:partialF}
\alpha'^3\, f^{(h,n)}(\tilde{s},\tilde{t})=\frac{1}{s^3}\sum^\infty_{\ell\text{ even}}\epsilon_\ell^{(h,n)} \,P_\ell(z)\,,
\end{equation} 
where for consistency with the tree-level expansion \eqref{eq:tree-level_exp} we have $\epsilon_\ell^{(0,n)}\equiv \epsilon_\ell^{(n)}$. Using these definitions, we can adapt the previously derived relation \eqref{ddiscLloops2} to the present context.\footnote{Taking into account the conventions for the genus expansion \eqref{eq:genus_exp}, this amounts to the replacements $a_\ell^{\text{Disc},L}\mapsto \kappa_{10}\R^4 g_s^{2h}\times \epsilon_\ell^{(h,n)}$, $a_\ell^{(0)}\mapsto \kappa_{10}\R^4\times\epsilon^{(n)}$ and $L\mapsto h$ in \eqref{ddiscLloops2}. Moreover, the factor of $L!$ cancels when \eqref{ddiscLloops2} is plugged into \eqref{eq:logL_coeff}.}
By matching the orders in $\ap$ on both sides, one finds that $\epsilon_\ell^{(h,n)}$ is determined by all possible distributions of $n$ into $h+1$ silos, i.e.
\begin{empheq}[box=\fbox]{equation}
\begin{aligned}\label{eq:StringLoops}
\epsilon_\ell^{(h,n)}= \left(-\frac{1}{4 \pi}\cdot\kappa_{10}^2\,s^4\right)^h \sum_{\sigma_{n,h}}\epsilon_\ell^{(n_1)}\epsilon_\ell^{(n_2)}\cdots \epsilon_\ell^{(n_{h{+}1})}\,,
\end{aligned}
\end{empheq}
where we iteratively used the self-replicating property \eqref{eq:sewing_relation} of the kinematical prefactor, giving rise to an extra factor of $\kappa_{10}^2s^4$ from every two-particle cut. Here $\sigma_{n,h}$ represents all possible solutions to $\sum_{i=1}^{h+1} n_i=n$, with all $n_i\geq0$.\footnote{The absence of terms at order $\ap$, $\ap^2$ and $\ap^4$ in the tree-level expansion \eqref{eq:A0_exp} implies that the corresponding partial-wave coefficients $\epsilon_\ell^{(n)}$ vanish, effectively giving that $n_i\neq1,2,4$.} Note that by summing over all permutations this formula automatically encodes the correct combinatorial factor coming from the various possible gluings of different vertices in the unitarity cut.

To recap, we have shown that the partial-wave coefficients $\epsilon_\ell^{(h,n)}$ for the $\alpha'$-expansion of the genus-$h$ leading log can be obtained from the tree-level coefficients $\epsilon_\ell^{(n)}$ in a purely algebraic fashion. The main formula for this is \eqref{eq:StringLoops}. Once the partial-wave coefficients are obtained, the full functions  $f^{(h,n)}(\tilde{s},\tilde{t})$ can be straightforwardly reconstructed via \eqref{eq:partialF}.\\\vspace{-0.2cm}

To illustrate the workings of the above main formula, let us explicitly spell out \eqref{eq:StringLoops} for the first few values of $n$, but keeping the genus $h$ general. Beginning with supergravity ($n=0$), the partial-wave coefficients of the maximal $s$-channel cut at genus $h$ read
\begin{align}\label{eq:eps_h0}
	\epsilon_\ell^{(h,0)} =\left(-\frac{1}{4 \pi}\cdot\kappa_{10}^2\,s^4\right)^h (\epsilon^{(0)}_\ell)^{h+1}\,,\quad \ell\in2\mathbb{N}\,.
\end{align}
Now, recall that the tree-level supergravity partial-wave coefficients $\epsilon_\ell^{(0)}$ from \eqref{sugrashifts} have infinite spin support, and as a consequence also the $\epsilon_\ell^{(h,0)}$ on the LHS have this feature. When plugged into \eqref{eq:partialF}, the corresponding coefficient functions $f^{(h,0)}(\stil,\ttil)$ are given in terms of an \textit{infinite} sum of partial waves. As we will see later on, these resummations give rise to functions with non-trivial transcendental weight.

On the other hand, higher-derivative corrections are spin truncated, c.f.~\eqref{treelshifts}, and the corresponding functions $f^{(h,n\geq3)}(\stil,\ttil)$ are therefore given by \emph{finite} sums of partial waves. For instance, the first string correction ($n=3$) has only support on $\ell=0$:
\begin{eqnarray}\label{eq:eps_h3}
\begin{split}
\epsilon_0^{(h,3)} &= \left(-\frac{1}{4 \pi}\cdot\kappa_{10}^2\,s^4\right)^h (h+1) \,(\epsilon_\ell^{(0)})^h \epsilon_0^{(3)}\,, &&\ell=0\,,\\ 
\epsilon_\ell^{(h,3)} &= 0\,, && \ell>0\,.
\end{split}
\end{eqnarray}
Similarly, for $n=5$ we have that only the first two spins are non-trivial:
\begin{eqnarray}\label{eq:eps_h5}
\begin{split}
\epsilon_\ell^{(h,5)} &= \left(-\frac{1}{4 \pi}\cdot\kappa_{10}^2\,s^4\right)^h (h+1) \,(\epsilon_\ell^{(0)})^h \epsilon_\ell^{(5)}\,, &&\ell = 0,2\,, \\ 
\epsilon_\ell^{(h,5)} &= 0\,, && \ell>2\,.
\end{split}
\end{eqnarray}
The next correction, with $n=6$, has still support on $\ell=0,2$ only, but is the first case where the sum in \eqref{eq:StringLoops} produces an additional contribution to the spin 0 partial wave:
\begin{eqnarray}\label{eq:eps_h6}
\begin{split}
\epsilon_0^{(h,6)} &=\left(-\frac{1}{4 \pi}\cdot\kappa_{10}^2\,s^4\right)^h (h+1)\bigg((\epsilon_0^{(0)})^h \epsilon_0^{(6)}+\frac{h}{2} (\epsilon_0^{(0)})^{h-1} (\epsilon_0^{(3)})^2\bigg), && \ell=0\,,\quad\\
\epsilon_2^{(h,6)} &= \left(-\frac{1}{4 \pi}\cdot\kappa_{10}^2\,s^4\right)^h (h+1) (\epsilon_0^{(0)})^h \epsilon_2^{(6)} \,, && \ell=2\,,\\
\epsilon_0^{(h,6)} &= 0\,, && \ell>2\,.
\end{split}
\end{eqnarray}
In general, the highest spin partial-wave coefficients (i.e. the leading Regge trajectory) is always the simplest, in the sense that is given by an insertion of a \textit{single} string correction in the cut. These leading trajectory partial-wave coefficients are given by
\begin{eqnarray}\label{singletrajec}
\begin{split}
\epsilon_{n-3}^{(h,n)} &= \left(-\frac{1}{4 \pi}\cdot\kappa_{10}^2\,s^4\right)^h (h+1) (\epsilon_{n-3}^{(0)})^h \epsilon_{n-3}^{(n)} \,,  && \quad n \in \{3,5,7,9,\ldots\}\,,\\ 
\epsilon_{n-4}^{(h,n)} &= \left(-\frac{1}{4 \pi}\cdot\kappa_{10}^2\,s^4\right)^h (h+1) (\epsilon_{n-4}^{(0)})^h \epsilon_{n-4}^{(n)} \,, && \quad n \in \{6,8,10,\ldots\}\,.
\end{split}
\end{eqnarray}
Next, we will put to use the general results for $\epsilon_\ell^{(h,n)}$ derived above. In particular, to illustrate the simplicity of our formalism, we will provide some explicit results for the leading log's up to genus 3.

%%%%%%%%%%%%%%%%%%%%%%%%%%%%%%%%%%%%%%%%%%%%%%%%%%%%%%%%%%%%%%%%%%%%%%%%%%%%%%%%%%%%%%%%%%
\subsection{Explicit results for leading log's}\label{sec:leading_logs_closed}
We start by introducing some useful notation. To represent which vertices $V_i$ enter the unitarity cut, we will schematically write $V_1\vert V_2\vert\cdots\vert V_k$, where the vertical bar $\vert$ stands for the $s$-channel unitarity cut, cf. Figure \ref{fig:s_cut}. The vertices $V_i$ can be either $S$, standing for supergravity and corresponding to an insertion of the partial-wave coefficients $\epsilon_\ell^{(0)}$, or tree-level higher-derivative terms $\partial^{2n-6}\R^4$, corresponding to the insertion of spin-truncated coefficients $\epsilon_\ell^{(n\geq3)}$.\footnote{For example, with this notation the gluing of $\epsilon_\ell^{(0)}$ and $\epsilon_\ell^{(5)}$ will be denoted by $S\vert\partial^4\R^4$. As can be read off from eqs. \eqref{eq:fh_exp}-\eqref{eq:StringLoops}, this term contributes at genus one and order $\ap^9$ with $f^{(1,5)}(\stil,\ttil)\log(-\stil)$.}
This notation is also employed in Table \ref{tab:double_exp}, where the more general structure of logarithmic terms in the low-energy expansion is summarised. However, for the moment we will only focus on the leading log's at each genus, i.e. $\log^h(-\stil)$ coming from the maximal number of cuts, which are coloured in red in Table \ref{tab:double_exp}.

%%%%%%%%%%%%%%%%%%%%%%%%%%%%%%%%%%%%%%
\subsubsection{Genus 1}\label{subsec:genus_1}
%%%%%%%%%%%%%%%%%%%%%%%%%%%%%%%%%%%%%%
At genus 1, the $s$-channel leading log is simply $\log(-\stil)$ with coefficient functions $f^{(1,n)}(\stil,\ttil)$. After adding up all crossing images, the $\ap$-expansion of the full amplitude takes the form
\begin{align}\label{eq:A1}
	A^{(1)} &= \ap^3 A^{(1,0)}+\ap^3\left(\sum_{n=3}^\infty f^{(1,n)}(\stil,\ttil)\log(-\stil)+f^{(1,n)}(\ttil,\stil)\log(-\ttil)+f^{(1,n)}(\util,\ttil)\log(-\util)\right)\notag\\[3pt]
	&\qquad\qquad\quad + (\texttt{analytic})\,,
\end{align}
where we have separated the non-analytic from the analytic ones.\footnote{It is important to note that the splitting between analytic and non-analytic terms is not unique, but does not affect the coefficient functions $f^{(1,n)}(\stil,\ttil)$ which our method has access to. For more details, see the proposal of ref. \cite{DHoker:2019blr} for a natural split into analytic and non-analytic contributions at genus 1.} We give more details on the $\ap$-expansion of the analytic terms in Section \ref{sec:general_log_structure}. Focussing on the non-analytic terms in the first line, the first contribution is the one-loop supergravity term $A^{(1,0)}$ which scales linearly with $\stil$ such that the total contribution is of order $\ap^4$. As mentioned earlier, computing its discontinuity $f^{(1,0)}(\stil,\ttil)$ requires some care as it involves an infinite sum over spins. Nevertheless, one can verify that the sum \eqref{eq:partialF} over partial waves with coefficients \eqref{eq:eps_h0} with $h=1$ resums to 
\begin{align}
\text{S}\vert\text{S}:\quad f^{(1,0)}(\tilde{s},\tilde{t}) = \frac{32 \pi^2}{15} \Big(\mathcal{B}(\tilde{s},\tilde{t})+\mathcal{B}(\tilde{s},\tilde{u}) \Big),
\end{align}
where $\mathcal{B}(s,t)$ is the $s$-channel discontinuity of the one-loop box integral in 10 dimensions given by
\begin{equation}
\mathcal{B}(\stil,\ttil) =  \frac{\stil^2}{\util^3} \Big(2\ttil^2 \log(-\stil/\ttil)  -\util(\stil+3\ttil)\Big).
\end{equation}

The subsequent terms in the first line of \eqref{eq:A1_log1} give the infinite tower of string corrections. As emphasised before, these contributions are much easier to determine since they are given by \textit{finite} sums over partial waves. 
For instance, plugging in the partial-wave coefficients from equations \eqref{eq:eps_h3}-\eqref{eq:eps_h6} with $h=1$, we obtain the next few orders in $\ap$,
\begin{align}
\begin{split}
	\text{S}\vert\R^4:\qquad f^{(1,3)}(\tilde{s},\tilde{t}) &= -\frac{2 \pi^2  \tilde{s}^4\zeta_3}{45}\,,\\
	\text{S}\vert\partial^4\R^4:\qquad	f^{(1,5)}(\tilde{s},\tilde{t}) &= -\frac{\pi^2  \tilde{s}^6 \zeta_5}{40320} (z^2+87)\,,\\
	\text{S}\vert\partial^6\R^4,\R^4\vert\R^4:\qquad	f^{(1,6)}(\tilde{s},\tilde{t}) &= \frac{\pi^2  \tilde{s}^7 \zeta_3^2}{161280} (z^2-49)\,.\hspace{2cm}
\end{split}
\end{align}
These are the $n=3,5,6$ terms in the sum \eqref{eq:A1}, and higher orders are computed straightforwardly. Note that in the last line there are two contributions to the $n=6$ term, one from the product of $\epsilon_\ell^{(0)}\times\epsilon_\ell^{(6)}$ and another one from $\epsilon_\ell^{(3)}\times\epsilon_\ell^{(3)}$. Moreover, there is an important relative factor of 2 between these two contributions, coming from the combinatorics encoded in formula \eqref{eq:StringLoops}.
Putting everything together, the low-energy expansion of the $s$-channel leading log at genus one is given by
\begin{align}\label{eq:A1_log1}
\begin{split}
A^{(1)}\vert_{\log(-\stil)}&= \ap^3\bigg(f^{(1,0)}(\tilde{s},\tilde{t}) - \frac{2\pi^2\zeta_3}{45}\,\tilde{s}^4 - \frac{\pi^2\zeta_5}{10080}\,\tilde{s}^4\big(22\tilde{s}^2-\tilde{t}\tilde{u}\big)\quad\\
&\qquad\quad-\frac{\pi^2\zeta_3^2}{40320}\,\tilde{s}^5\big(12\tilde{s}^2+\tilde{t}\tilde{u}\big)\\
&\qquad\quad-\frac{\pi^2\zeta_7}{2419200}\stil^4\big(260\stil^4-25\stil^2\ttil\util+2\ttil^2\util^2\big)\\
&\qquad\quad-\frac{\pi^2\zeta_3\zeta_5}{2419200}\stil^5\big(70\stil^4+5\stil^2\ttil\util-\ttil^2\util^2\big)\bigg)+\o(\ap^{13})\,,
\end{split}
\end{align}
where we have added further contributions from $S\vert\partial^8\R^4$ in the third line, and the combined sum from $S\vert\partial^{10}\R^4$ and $\R^4\vert\partial^4\R^4$ in the fourth line. The above expansion is in agreement with known results in the literature \cite{DHoker:2019blr,Edison:2021ebi,Eberhardt:2022zay}, see also \cite{Green:1999pv,Green:2008uj} for earlier works.

Let us stress that using our formalism all we had to do is to multiply the tree-level partial-wave coefficients and then resum them against finitely many partial waves. Obtaining higher orders in $\ap$ is therefore straightforward and can be implemented very efficiently.

%%%%%%%%%%%%%%%%%%%%%%%%%%%%%%%%%%%%%%
\subsubsection{Genus 2}\label{subsec:genus_2}
%%%%%%%%%%%%%%%%%%%%%%%%%%%%%%%%%%%%%%
The $s$-channel leading log at genus 2 is now given by $\log^2(-\stil)$, arising from the insertion of \textit{three} tree-level partial-wave coefficients into the unitarity cut. In analogy with the genus-one amplitude, the crossing completion of the leading log implies that the low-energy expansion of the full genus-two amplitude $A^{(2)}$ has the structure
\begin{align}\label{eq:A2}
	A^{(2)} &= \ap^3 A^{(2,0)}+ \ap^3\left(\sum_{n=3}^\infty f^{(2,n)}(\stil,\ttil)\log^2(-\stil)+f^{(2,n)}(\ttil,\stil)\log^2(-\ttil)+f^{(2,n)}(\util,\ttil)\log^2(-\util)\right)\notag\\[3pt]
	&\qquad\qquad\quad + (\texttt{lower log's}) + (\texttt{analytic})\,,
\end{align}
where in contrast to the genus-one case there will be further sub-leading non-analytic terms which do contribute to the \textit{single} discontinuity, schematically denoted by \texttt{lower log's} above. It is important to point out that this includes not only single log's, but generically also double log's of the form $\log(-\stil)\log(-\ttil)$, $\log(-\stil)\log(-\util)$ etc., which do not contribute to the double discontinuity in any channel. Since our method relies on leveraging iterated two-particle $s$-channel cuts, we cannot access such terms.

The first term in the above sum is given by the two-loop supergravity amplitude $A^{(2,0)}$, see e.g. \cite{Bern:1998ug,Bissi:2020woe}. We have only access to its $\log^2(-\stil)$ coefficient (denoted by $f^{(2,0)}$), however. It arises from the insertion $S\vert S\vert S$ into the unitarity cut relation, and contributes at order $\ap^8$ with
\begin{align}
\begin{split}
	f^{(2,0)}(\stil,\ttil) &= \frac{16\pi^4\stil^6}{3375\ttil^3\util^3}\Big[30\stil(21\stil^4-75\stil^2\ttil\util+25\ttil^2\util^2)\log^2(-\stil/\util)\\
	&\qquad\qquad\quad -\big(\stil(2\stil^4+1235\stil^2\ttil\util-2435\ttil^2\util^2)-60\util^3(15\stil^2-5\ttil^2+6\util^2)\log(\ttil/\util)\big)\log(-\stil/\util)\\
	&\qquad\qquad\quad+\ttil\big(-10\pi^2\ttil^2(10\stil^2-10\stil\ttil+\ttil^2)+\stil(1258\stil^3+4161\stil^2\ttil+5806\stil\ttil^2+2903\ttil^3)\\
	&\qquad\qquad\qquad\quad-(1260\stil^4+3870\stil^3\ttil+4170\stil^2\ttil^2+1785\stil\ttil^3+227\ttil^4)\log(\ttil/\util)\big)\\
	&\qquad\qquad\quad-60(\ttil-\util)(21\stil^4-33\stil^2\ttil\util+\ttil^2\util^2)\text{Li}_2(-\ttil/\util)\Big],
\end{split}
\end{align}
which agrees with the result of \cite{Bissi:2020woe} up to an overall normalisation.\footnote{The detailed comparison with the expression given in equations (3.11) and (B.8) of \cite{Bissi:2020woe} works as follows: To match against our partial-wave resummation, one has to analytically continue their expression so that all logarithms are real in the physical region. Moreover, we found that their result represents in fact the $t$-channel double-cut (and not the $s$-channel one, as suggested by their (3.11)), i.e. the coefficient of $\log^2(-\ttil)$. To recover the $s$-channel cut one needs to perform an $\stil\leftrightarrow\ttil$ exchange.}
As before, the subsequent string corrections are straightforward to compute. Here we skip their detailed derivation and simply quote our result for the low-energy expansion of the genus-two $\log^2(-\stil)$ term:
\begin{align}\label{eq:A2_log2}
\begin{split}
A^{(2)}\vert_{\log^2(-\stil)}&=\ap^3\bigg(f^{(2,0)}(\tilde{s},\tilde{t}) + \frac{\pi^4\zeta_3}{1350}\,\tilde{s}^8 + \frac{\pi^4\zeta_5}{2^9\cdot 33075}\,\tilde{s}^8\big(610\tilde{s}^2-\tilde{t}\tilde{u}\big)\\
&\qquad\quad+\frac{\pi^4\zeta_3^2}{2^{11}\cdot 33075}\,\tilde{s}^9\big(510\tilde{s}^2+\tilde{t}\tilde{u}\big)\\
&\qquad\quad+\frac{\pi^4\zeta_7}{2^{13}\cdot7441875}\stil^8\big(106990\stil^4-365\stil^2\ttil\util+4\ttil^2\util^2\big)\\
&\qquad\quad+\frac{\pi^4\zeta_3\zeta_5}{2^{11}\cdot7441875}\stil^9\big(22265\stil^4+35\stil^2\ttil\util-\ttil^2\util^2\big)\bigg)+\o(\ap^{17})\,.
\end{split}
\end{align}
This constitutes a new result, which we hope can be matched by future worldsheet computations. We stress again that computing higher orders can be implemented very easily. 

%%%%%%%%%%%%%%%%%%%%%%%%%%%%%%%%%%%%%%
\subsubsection{Genus 3}\label{subsec:genus_3}
%%%%%%%%%%%%%%%%%%%%%%%%%%%%%%%%%%%%%%
Lastly, we consider the genus-three contribution $A^{(3)}$. The unitarity cuts predict the following structure for the amplitude:
\begin{align}\label{eq:A3}
	A^{(3)} &=\ap^3 A^{(3,0)}+\ap^3\left(\sum_{n=3}^\infty f^{(3,n)}(\stil,\ttil)\log^3(-\stil)+f^{(3,n)}(\ttil,\stil)\log^3(-\ttil)+f^{(3,n)}(\util,\ttil)\log^3(-\util)\right)\notag\\[3pt]
	&\qquad\qquad\quad + (\texttt{lower log's}) + (\texttt{analytic})\,,
\end{align}
where by $\texttt{lower log's}$ we denote terms which do not have a triple discontinuity in any channel. The three-loop supergravity amplitude\footnote{See e.g. \cite{Bern:2007hh,Bern:2008pv,Bern:2010ue} for details on the structure of its integrand and UV properties of the amplitude.} $A^{(3,0)}$ scales as $\stil^9$, such that the low-energy expansion of the triple $s$-channel cut starts at order $\ap^{12}$. For the first few terms we obtain the expansion
\begin{align}\label{eq:A3_log3}
\begin{split}
A^{(3)}\vert_{\log^3(-\stil)}&=\ap^3\bigg(f^{(3,0)}(\tilde{s},\tilde{t}) - \frac{\pi^6\zeta_3}{91125}\,\tilde{s}^{12} - \frac{\pi^6\zeta_5}{2^{10}\cdot31255875}\,\tilde{s}^{12}\big(17074\tilde{s}^2-\tilde{t}\tilde{u}\big)\\
&\qquad\quad-\frac{\pi^6\zeta_3^2}{2^{12}\cdot31255875}\,\tilde{s}^{13}\big(18990\tilde{s}^2+\tilde{t}\tilde{u}\big)\\
&\qquad\quad-\frac{\pi^6\zeta_7}{2^{14}\cdot105488578125}\stil^{12}\big(44903030\stil^4-5455\stil^2\ttil\util+8\ttil^2\util^2\big)\\
&\qquad\quad-\frac{\pi^6\zeta_3\zeta_5}{2^{12}\cdot105488578125}\stil^{13}\big(6220715\stil^4+260\stil^2\ttil\util-\ttil^2\util^2\big)\bigg)\\
&\quad+\o(\ap^{21})\,.
\end{split}
\end{align}

%%%%%%%%%%%%%%%%%%%%%%%%%%%%%%%%%%%%%%%%%%%%%%%%%%%%%%%%%%%%%%%%%%%%%%%%%%%%%%%%%%%%%%%%%%
\subsection{All-loop resummation of the leading Regge trajectory}\label{sec:leading_trajectory}
As commented around equation \eqref{singletrajec}, the partial-wave coefficients of the leading Regge trajectory take a very simple form, as they contain a single string correction. In particular, at fixed $n$ the highest-spin partial wave coefficient (i.e. $\ell=n{-}3$ for $n$ odd, and $\ell=n{-}4$ for $n$ even) at genus $h$ is given by the string tree-level partial-wave coefficient $\epsilon^{(n)}_\ell$ multiplied by $h$ powers of the supergravity $\epsilon^{(0)}_\ell$. In the amplitude, this implies that the corresponding $\log$'s can be easily resumed. For example, for odd $n\geq3$, the resummation of $\log$'s leads to 
\begin{equation}
\sum_{h=0}^\infty \big[f^{(h,n)}(\stil,\ttil)\vert_{\tilde{t}^{n-3}}\big]\cdot\log^h(-\tilde{s}) = \frac{1}{8^{n-3}}\,\frac{2\,\zeta_{n}}{\Big(1 + \frac{8 \pi^2 \tilde{s}^4\log(-\tilde{s})}{(n-2)_6}\Big)^2}\,,
\end{equation}
where we focus on the leading power of $\ttil$ of the partial waves, given by $\ttil^{n-3}$. For even $n$ instead, the resummation yields
\begin{equation}
\sum_{h=0}^\infty \big[f^{(h,n)}(\stil,\ttil)\vert_{\tilde{t}^{n-4}}\big]\cdot\log^h(-\tilde{s}) = -\frac{4}{8^{n-3}}\bigg(\sum_{i=1}^{\frac{n-4}{2}} \zeta_{2i+1}\zeta_{n-2i-1} \bigg) \, \frac{\tilde{s}}{\Big( 1 + \frac{8 \pi^2 \tilde{s}^4\log(-\tilde{s})}{(n-3)_6}\Big)^2} \,.
\end{equation}
This can be interpreted as the solution to the RG equation for the leading trajectory operators, which only involves supergravity contributions. 

%%%%%%%%%%%%%%%%%%%%%%%%%%%%%%%%%%%%%%%%%%%%%%%%%%%%%%%%%%%%%%%%%%%%%%%%%%%%%%%%%%%%%%%%%%
\section{The general structure of logarithmic terms}\label{sec:general_log_structure}
%%%%%%%%%%%%%%%%%%%%%%%%%%%%%%%%%%%%%%%%%%%%%%%%%%%%%%%%%%%%%%%%%%%%%%%%%%%%%%%%%%%%%%%%%%
So far, we have seen how products of tree-level partial-wave coefficients compute the leading logarithmic terms in the closed string amplitude. As we will discuss now, the partial-wave expansion can actually teach us more than that: for certain special orders in $g_s$ and $\ap$ which do not have a leading log, we can also access logarithmic terms of lower degree.

The general mechanism is in fact in complete analogy with the previous discussion: any further analytic term beyond tree-level (i.e. a term which is not already superseded by some logarithmic contribution) will feed into the unitarity cut and give rise to new non-analytic terms, which consequently are of lower degree than the maximal log. Very neatly, this mechanism thus predicts the general structure of logarithmic terms in the low-energy expansion, which we summarise in Table \ref{tab:double_exp} below.\\\vspace{-0.2cm}

Let us first illustrate how this works with a concrete example. Consider the genus-one amplitude $A^{(1)}$, whose $\ap$-expansion starts with an analytic term at order $\ap^3$ (one order before the non-analytic one-loop supergravity contribution $S\vert S$). This term is in fact the genus-one completion of the tree-level $\R^4$ term, which we denote by $\R^4_1$. Inserting this vertex into the unitarity cut together with one tree-level supergravity vertex, i.e. $S\vert\R^4_1$ for short, yields a single $\log(-\stil)$ contribution to the genus-two amplitude at order $\ap^7$. In the low-energy expansion of $A^{(2)}$, this term will thus appear one order \textit{before} the leading log contribution of the two-loop supergravity term  $S\vert S\vert S$.

In a similar fashion, inserting the genus-one $\R^4_1$ vertex \textit{twice} into the unitarity cut gives a single $\log(\stil)$ contribution $\R^4_1\vert\R^4_1$ to the genus-\textit{three} amplitude $A^{(3)}$ at order $\ap^{10}$. In fact, it so happens that there is another single $\log(\stil)$ contribution at the same order, coming from the insertion $S\vert\partial^6\R^4_2$, induced by the non-vanishing genus-two correction to the tree-level $\partial^6\R^4$ term.

More generally, inserting two analytic terms at genus $h_1$, $h_2$ and $\ap$-orders $\ap^{k_1}$, $\ap^{k_2}$ into the unitarity cut gives rise to a single $\log(-\stil)$ non-analytic term at genus $h'=h_1+h_2+1$ and order $k'=k_1+k_2+4$ in $\ap$. The generalisation to more insertions, and therefore higher log's, proceeds in complete analogy to the leading log.\footnote{
In full generality, feeding a number of $p$ various contributions from analytic terms at genera $h_i$ and $\ap$-orders $k_i$ ($i=1,\ldots,p$) into the unitarity cut yields a $\log^{p-1}(-\stil)$ non-analytic term at genus $h'=p-1+\sum_{i=1}^p h_i$ and $\ap$-order $n'=\sum_{i=1}^p n_i$.}
In particular, the precise coefficient functions can again be computed from the partial-wave decomposition by a simple adaptation of formulae \eqref{eq:partialF} and \eqref{eq:StringLoops}.\\\vspace{-0.2cm}

By listing all non-vanishing analytic terms at higher-genus and iterating this mechanism, we arrive at an interesting structure of logarithmic terms in the low-energy expansion. Up to genus 6, this is visualised in Table \ref{tab:double_exp}. For high enough genus, the generic pattern is as follows: For the first few orders in $\ap$ only analytic terms can appear. Then, for $h\geq5$, the first non-analytic term which can appear is a single log at order $\ap^{11}$, sourced by the genus-$(h-1)$ corrections to the tree-level $\partial^8\R^4$ term. Note that their precise coefficients are as of now unknown, but it is our understanding that they are generally expected to be non-vanishing. Starting from order $\ap^{15}$ a double log can appear, and so on for higher logs, until the maximum logarithmic degree is reached by the first leading log at order $\ap^{4h}$, which is given by the $h$-loop supergravity contribution.

In the remainder of this section, we put these ideas to use and consider the low-energy expansion of the genus-two and -three amplitudes. In order to give some quantitative predictions, we first reproduce the coefficients of currently known higher-genus analytic terms. From these, we then extract the partial-wave coefficients and feed them into the unitarity cut as seen in previous sections.

%%%%%%%%%%%%%%%%%%%%%%%%%%%%%%%%%%%%%%%%%%%%%%%%%%%%%%%%%%%%%%%%%%%%%%%%%%%%%%%%%%%%%%%%%%
\subsection{Collection of known analytic terms at higher genus}\label{subsec:known_analytic_terms}
Almost all information about higher-genus analytic terms concerns the first three higher-derivative corrections, namely $\R^4$, $\partial^4\R^4$ and $\partial^6\R^4$. Preserving some amount of supersymmetry, they only have a \textit{finite} number of higher-genus corrections: the $\R^4$ term receives contributions up to genus one, the $\partial^4\R^4$ term up to genus two (with the notable feature that the genus-one coefficient vanishes), and lastly the $\partial^6\R^4$ term up to genus three. On the other hand, it is our understanding that the unprotected terms at higher order in $\ap$ (starting from the $\partial^8\R^4$ term) will generically have an infinite number of higher-genus corrections (see e.g. reference \cite{Bjornsson:2010wm}). This is reflected in Table \ref{tab:double_exp} by the vertical double-line, which separates the protected and unprotected higher-derivative corrections.

Let us now compile the numerical values of the known analytic contributions. In our conventions, the genus-one terms read \cite{Green:1999pv,Green:2008uj,DHoker:2015gmr,DHoker:2019blr}
\begin{align}\label{eq:higher1}
	A^{(1)}\vert_{\R^4_1} = \ap^3\,\frac{2\pi^2}{3}\,,\quad A^{(1)}\vert_{\partial^6\R^4_1} = \ap^3\,\frac{\pi^2\zeta_3}{288}\,\sigma_3\,,\quad A^{(1)}\vert_{\partial^{10}\R^4_1} = \ap^3\,\frac{29\pi^2\zeta_5}{2^{11}\cdot135}\,\sigma_2\sigma_3\,,
\end{align}
where we included the known contribution from the unprotected $\partial^{10}\R^4_1$ term. Note that the $\partial^4\R^4_1$ term is not listed since its coefficient vanishes. The $\partial^8\R^4_1$ correction, as well as all terms starting from $\partial^{12}\R^4_1$, are also omitted since they are superseded by the presence of non-analytic terms at those orders.\footnote{The full genus-one correction to the $\partial^8\R^4$ and $\partial^{12}\R^4$ terms, including both analytic and non-analytic contributions, has been worked out in \cite{DHoker:2019blr}.}

Beyond genus one, the only known analytic terms are the higher-genus completions of the protected terms mentioned earlier. The genus-two terms read \cite{DHoker:2005jhf,Green:2005ba,DHoker:2014oxd}
\begin{align}\label{eq:higher2}
	A^{(2)}\vert_{\partial^4\R^4_2} = \ap^3\,\frac{\pi^4}{1080}\sigma_2\,,\quad A^{(2)}\vert_{\partial^6\R^4_2} = \ap^3\,\frac{\pi^4}{1440}\,\sigma_3\,,\quad A^{(2)}\vert_{\partial^{12}\R^4_2} = \ap^3\,\pi^4\big(D_1\sigma_2^3+D_2\sigma_3^2\big)\,,
\end{align}
and we parametrised the unprotected $\partial^{12}\R^4_2$ correction with some currently unknown coefficients $D_1$ and $D_2$. Interestingly, assuming uniform transcendentality of the sub-leading log's will lead us to conjecture that these coefficients are proportional to $\zeta_3$. This will be explained in more detail in Section \ref{subsec:transcendentality}. 

Finally, at genus three, the only known coefficient is the leading order term \cite{Green:2005ba,Gomez:2013sla}
\begin{align}\label{eq:higher3}
	A^{(3)}\vert_{\partial^6\R^4_3} = \ap^3\,\frac{\pi^6}{408240}\,\sigma_3\,.
\end{align}
The partial-wave coefficients of these terms are easily extracted using e.g. equation \eqref{eq:pw_monomial}. For reference, these expressions are recorded in Appendix \ref{app:pw_coeffs}.

%%%%%%%%%%%%%%%%%%%%%%%%%%%%%%%%%%%%%%%%%%%%%%%%%%%%%%%%%%%%%%%%%%%%%%%%%%%%%%%%%%%%%%%%%%
\definecolor{forestgreen(web)}{rgb}{0.13, 0.55, 0.13}
\definecolor{candyapplered}{rgb}{1.0, 0.03, 0.0}
\definecolor{deepskyblue}{rgb}{0.0, 0.75, 1.0}
\renewcommand\analytic[1]{{\color{forestgreen(web)}#1}}
\renewcommand\leading[1]{{\color{candyapplered}#1}}
\renewcommand\sleading[1]{{\color{deepskyblue}#1}}
\begin{landscape}
\begin{table}
\begin{center}
\def\cwidth{1.3cm}
\def\cwidthalt{0.8cm}
\def\cwidthaltt{0.4cm}
$\begin{array}{|c|>{$}C{\cwidthaltt}<{$}|>{$}C{\cwidthaltt}<{$}|>{$}C{\cwidthaltt}<{$}|>{$}C{\cwidthalt}<{$}|>{$}C{\cwidthalt}<{$}|>{$}C{\cwidthalt}<{$}|>{$}C{\cwidthalt}<{$}||>{$}C{\cwidth}<{$}|>{$}C{\cwidth}<{$}|>{$}C{\cwidth}<{$}|>{$}C{\cwidth}<{$}|>{$}C{\cwidth}<{$}|>{$}C{\cwidth}<{$}|>{$}C{\cwidth}<{$}|>{$}C{\cwidth}<{$}|>{$}C{\cwidth}<{$}|>{$}C{\cwidth}<{$}|>{$}C{\cwidthalt}<{$}}\hline
{(\ap)^n} & 0 & 1 & 2 & 3 & 4 & 5 & 6 & 7 & 8 & 9 & 10 & 11 & 12 & 13 & 14 & 15 & 16 & \cdots \\ \hline
&&&&&&&&&&&&&&&&&&\\[-10pt]
\,\text{genus 0}~ & \analytic{S} &  &  & \analytic{\R_0} &  & \analytic{_4\R_0} & \analytic{_6\R_0} & \analytic{_8\R_0} & \analytic{_{10}\R_0} & \analytic{_{12}\R_0} & \analytic{_{14}\R_0} & \analytic{_{16}\R_0} & \analytic{_{18}\R_0} & \analytic{_{20}\R_0} & \analytic{_{22}\R_0} & \analytic{_{24}\R_0} & \analytic{_{26}\R_0} & \cdots \\[4pt] \hline
\text{genus 1} &  &  &  & \analytic{\R_1} & \leading{S\vert S} & \analytic{0} & \analytic{_6\R_1} & \leading{S\vert\R_0} & \analytic{_{10}\R_1} & \leading{S\vert_4\R_0} & \leading{\begin{array}{@{}c@{}}S\vert_6\R_0 \\ \R_0\vert\R_0\end{array}} & \leading{S\vert_8\R_0} & \leading{\begin{array}{@{}c@{}}S\vert_{10}\R_0 \\ \R_0\vert_4\R_0\end{array}} & \leading{\begin{array}{@{}c@{}}S\vert_{12}\R_0 \\ \R_0\vert_6\R_0\end{array}} & \leading{\scalebox{0.9}{$\begin{array}{@{}c@{}}S\vert_{14}\R_0 \\ \R_0\vert_8\R_0 \\ _4\R_0\vert_4\R_0\end{array}$}} & \leading{\scalebox{0.9}{$\begin{array}{@{}c@{}}S\vert_{16}\R_0 \\ \R_0\vert_{10}\R_0 \\ _4\R_0\vert_6\R_0\end{array}$}} & \leading{\scalebox{0.85}{\hspace{-0.05cm}$\begin{array}{@{}c@{}}S\vert_{18}\R_0 \\ \R_0\vert_{12}\R_0 \\ _4\R_0\vert_8\R_0 \\ _6\R_0\vert_6\R_0\end{array}$}} & \cdots \\ \hline
\text{genus 2} &  &  &  &  &  & \analytic{_4\R_2} & \analytic{_6\R_2} & \sleading{S\vert\R_1} & \leading{S\vert S\vert S} & _{12}\R_2 & \sleading{\begin{array}{@{}c@{}}S\vert_6\R_1 \\ \R_0\vert\R_1\end{array}} & \leading{\hspace{-0.05cm}S\vert S\vert\R_0} & \sleading{\begin{array}{@{}c@{}} S\vert_{10}\R_1 \\  _4\R_0\vert\R_1\end{array}} & \leading{\scalebox{0.95}{$S\vert S\vert_4\R_0$}} & \leading{\scalebox{0.9}{$\begin{array}{@{}c@{}}S\vert S\vert_6\R_0 \\ S\vert\R_0\vert\R_0\end{array}$}} & \leading{\scalebox{0.95}{$S\vert S\vert_8\R_0$}} & \leading{\scalebox{0.85}{\hspace{-0.04cm}$\begin{array}{@{}c@{}}S\vert S\vert_{10}\R_0 \\ S\vert\R_0\vert_4\R_0\end{array}$}} & \cdots \\ \hline
\text{genus 3} &  &  &  &  &  &  & \analytic{_6\R_3} & _8\R_3 & _{10}\R_3 & S\vert_4\R_2 & \begin{array}{@{}c@{}}S\vert_6\R_2 \\ \R_1\vert\R_1\end{array} & \sleading{S\vert S\vert\R_1} & \leading{\hspace{-0.04cm}S\vert S\vert S\vert S} & \scalebox{0.95}{$\begin{array}{@{}c@{}}S\vert_{12}\R_2 \\ \R_0\vert_6\R_2 \\ \R_1\vert_6\R_1\end{array}$} & \sleading{\scalebox{0.9}{$\begin{array}{@{}c@{}}S\vert S\vert_6\R_1 \\ S\vert\R_0\vert\R_1\end{array}$}} & \leading{\scalebox{0.85}{\hspace{-0.03cm}$S\vert S\vert S\vert\R_0$}} & \sleading{\scalebox{0.85}{$\begin{array}{@{}c@{}} S\vert S\vert_{10}\R_1 \\ S\vert_4\R_0\vert\R_1\end{array}$}} & \cdots \\ \hline
\text{genus 4} &  &  &  &  &  &  &  & _8\R_4 & _{10}\R_4 & _{12}\R_4 & S\vert_6\R_3 & S\vert_8\R_3 & \begin{array}{@{}c@{}} S\vert_{10}\R_3 \\ \R_1\vert_4\R_2 \end{array} & \scalebox{0.95}{$S\vert S\vert_4\R_2$} & \scalebox{0.9}{$\begin{array}{@{}c@{}} S\vert S\vert_{6}\R_2 \\ S\vert\R_1\vert\R_1 \end{array}$} & \sleading{\scalebox{0.85}{\hspace{-0.03cm}$S\vert S\vert S\vert\R_1$}} & \leading{\scalebox{0.8}{\hspace{-0.04cm}$S\vert S\vert S\vert S\vert S$}} & \cdots \\[7pt] \hline
\text{genus 5} &  &  &  &  &  &  &  & _8\R_5 & _{10}\R_5 & _{12}\R_5 & _{14}\R_5 & S\vert_8\R_4 & \hspace{-0.05cm}S\vert_{10}\R_4 & \hspace{-0.05cm}\begin{array}{@{}c@{}} S\vert_{12}\R_4 \\ \R_1\vert_6\R_3 \end{array} & \scalebox{0.95}{$S\vert S\vert_6\R_3$} & \scalebox{0.95}{$S\vert S\vert_8\R_3$} & \scalebox{0.85}{\hspace{-0.04cm}$\begin{array}{@{}c@{}} S\vert S\vert_{10}\R_3 \\ S\vert\R_1\vert_4\R_2 \end{array}$} & \cdots \\[7pt] \hline
\text{genus 6} &  &  &  &  &  &  &  & _8\R_6 & _{10}\R_6 & _{12}\R_6 & _{14}\R_6 & S\vert_8\R_5 & \hspace{-0.05cm}S\vert_{10}\R_5 & \hspace{-0.05cm}S\vert_{12}\R_5 & \begin{array}{@{}c@{}} S\vert_{14}\R_5 \\ \R_0\vert_8\R_5 \\ \R_1\vert_8\R_4 \end{array} & \scalebox{0.95}{$S\vert S\vert_8\R_4$} & \scalebox{0.9}{\hspace{-0.04cm}$S\vert S\vert_{10}\R_4$} & \cdots \\[16pt] \hline
\vdots &  &  &  &  &  &  &  & \vdots & \vdots & \vdots & \vdots & \vdots & \vdots & \vdots & \vdots & \vdots & \vdots & \ddots  \\
\end{array}$
\end{center}
\caption{Summary of analytic and non-analytic terms in the low-energy expansion of $A^{(h)}$ up to genus 6. At any given order in $g_s$ and $\ap$, we only show terms which contribute to the maximal $s$-channel cut. In case there is no non-analytic term at that order, we show the corresponding analytic contribution. \analytic{Known analytic terms} are highlighted in green. For non-analytic terms, \leading{leading log's} are coloured in {red}, \sleading{sub-leading log's} in {blue}.\\
To avoid cluttering, we have simplified the notation from the main text. Here $_a\R_b$ stands for $\partial^{a}\R^4_b$, denoting the genus-$b$ contribution to the $\partial^{a}\R^4$ higher derivative correction, with the understanding that $\R_b\equiv\,_0\R_b$.\\
Finally, a vertical double-line separates the protected (to the left) from the unprotected (to the right) higher-derivative terms. The protected ones receive only a finite number of higher-genus corrections, while the unprotected ones are expected to receive infinitely many corrections.}\label{tab:double_exp}
\end{table}
\end{landscape}
%%%%%%%%%%%%%%%%%%%%%%%%%%%%%%%%%%%%%%%%%%%%%%%%%%%%%%%%%%%%%%%%%%%%%%%%%%%%%%%%%%%%%%%%%%

%%%%%%%%%%%%%%%%%%%%%%%%%%%%%%%%%%%%%%%%%%%%%%%%%%%%%%%%%%%%%%%%%%%%%%%%%%%%%%%%%%%%%%%%%%
\subsection{The full low-energy expansion at higher genus}\label{subsec:sub_leading_logs}
The first instance where sub-leading logarithmic terms occur is at genus 2. The leading log's are given by $\log^2(-\stil)$, so there is room for sub-leading logarithmic terms $\log(-\stil)$. 

%%%%%%%%%%%%%%%%%%%%%%%%%%%%%%%%%%%%%%
\subsubsection{Genus 2}
%%%%%%%%%%%%%%%%%%%%%%%%%%%%%%%%%%%%%%
Let us now turn to the low-energy expansion of the genus-two amplitude $A^{(2)}$ in some more detail. For the first few orders in $\ap$, we find the following terms: the expansion starts with two analytic contributions at orders $\ap^5$ and $\ap^6$. These correspond to the $\partial^4\R^4_2$ and $\partial^6\R^4_2$ terms, whose values are given above in \eqref{eq:higher2}. Then, at order $\ap^7$, we encounter the first non-analytic contribution which is in fact a \textit{sub}-leading log coming from the gluing $S\vert\R^4_1$. Only at the next order, $\ap^8$, we arrive at the first leading log, given as usual by the supergravity contribution $S\vert S\vert S$. The logarithmic structure at higher orders can be easily read off from Table \ref{tab:double_exp}.

As described earlier, also the sub-leading log's can be computed using the partial-wave decomposition. Using as input the known analytic terms at genus one and two, c.f. Section \ref{subsec:known_analytic_terms}, we obtain the expansion
\begin{align}\label{eq:genus_2_final}
\begin{split}
	A^{(2)} &= \ap^3\bigg(\frac{\pi^4}{1080}\sigma_2 + \frac{\pi^4}{1440}\,\sigma_3 \\ &\quad\qquad-\bigg[\frac{2\pi^4}{135}\Big(\stil^4\log(-\stil)+\ttil^4\log(-\ttil)+\util^4\log(-\util)\Big)+(\texttt{analytic})\bigg] \\
	&\quad\qquad+A^{(2,0)}(\stil,\ttil)+\bigg[\pi^4\big(D_1\sigma_2^3+D_2\sigma_3^2\big)\bigg] \\
	&\quad\qquad-\bigg[\frac{\pi^4\zeta_3}{2^7\cdot945}\Big(\stil^5(18\stil^2+\ttil\util)\log(-\stil)+(\stil\leftrightarrow\ttil)+(\stil\leftrightarrow\util)\Big)+(\texttt{analytic})\bigg] \\
	&\quad\qquad+\bigg[\frac{\pi^4\zeta_3}{1350}\Big(\stil^8\log^2(-\stil)+\ttil^8\log^2(-\ttil)+\util^8\log^2(-\util)\Big)+(\texttt{lower log's})\bigg] \\
	&\quad\qquad-\bigg[\frac{\pi^4\zeta_5}{2^{11}\cdot212625}\Big(\stil^5(3115\stil^4+145\stil^2\ttil\util-29\ttil^2\util^2)\log(-\stil)\\[-10pt]
	&\qquad\qquad\qquad\qquad\qquad\qquad\qquad\qquad\qquad~+(\stil\leftrightarrow\ttil)+(\stil\leftrightarrow\util)\Big)+(\texttt{analytic})\bigg]\bigg)\\
	&\quad+\o(\ap^{13})\,.
\end{split}
\end{align}
All subsequent terms have a leading log, and as such have been discussed in Section \ref{subsec:genus_2}.

The above result constitutes a concrete prediction for the $\ap$-expansion at genus two. A first target for future worldsheet computations might be the sub-leading log at order $\ap^7$. It appears one order before the two-loop supergravity term $A^{(2,0)}(\stil,\ttil)$ and thus precedes the first genuine two-loop contribution, so in some sense it is purely a genus-one effect.

%%%%%%%%%%%%%%%%%%%%%%%%%%%%%%%%%%%%%%
\subsubsection{Genus 3}
%%%%%%%%%%%%%%%%%%%%%%%%%%%%%%%%%%%%%%
The analytic structure of the genus-three amplitude $A^{(3)}$ is even more intricate. The leading log's are given by $\log^3(-\stil)$, allowing for the presence of sub- and sub-sub-leading logarithmic terms.

So far, the only known coefficient in the low-energy expansion of $A^{(3)}$ is the first term at order $\ap^6$, given by the genus-three contribution to the $\partial^6\R^4$ higher-derivative correction, see equation \eqref{eq:higher3}. It is followed by two more analytic terms, denoted by $\partial^8\R^4_3$ and $\partial^{10}\R^4_3$ in the equation below, whose coefficients are unknown. The non-analytic terms which our method has access to start from order $\ap^9$. In particular, the next two orders are given by sub-sub-leading log's coming from $S\vert\partial^4\R^4_2$ at order $\ap^9$, and $S\vert\partial^6\R^4_2$ and $\R^4_1\vert\R^4_1$ at order $\ap^{10}$, respectively. They are followed by a sub-leading log at order $\ap^{11}$ obtained from $S\vert S\vert\R^4_1$, and only at order $\ap^{12}$ one arrives at a leading log contribution coming from the three-loop supergravity term, $S\vert S\vert S\vert S$. For higher-order terms we again refer to Table \ref{tab:double_exp}.

Putting everything together, the unitarity cut method predicts that the low-energy expansion of $A^{(3)}$ takes the form
\begin{align}\label{eq:genus_3_final}
\begin{split}
	A^{(3)} &= \ap^3\bigg(\frac{\pi^6}{408240}\,\sigma_3 +\big[\,\partial^8\R^4_3\propto\sigma_2^2\,\big]+\big[\,\partial^{10}\R^4_3\propto\sigma_2\sigma_3\,\big] \\ &\quad\qquad-\bigg[\frac{\pi^6}{2^4\cdot42525}\Big(\stil^4(22\stil^2-\ttil\util)\log(-\stil)+(\stil\leftrightarrow\ttil)+(\stil\leftrightarrow\util)\Big)+(\texttt{analytic})\bigg] \\
	&\quad\qquad-\bigg[\frac{\pi^6}{2^7\cdot4725}\Big(\stil^5(16\stil^2+\ttil\util)\log(-\stil)+(\stil\leftrightarrow\ttil)+(\stil\leftrightarrow\util)\Big)+(\texttt{analytic})\bigg] \\
	&\quad\qquad+\bigg[\frac{\pi^6}{4050}\Big(\stil^8\log^2(-\stil)+\ttil^8\log^2(-\ttil)+\util^8\log^2(-\util)\Big)+(\texttt{lower log's})\bigg] \\
	&\quad\qquad+A^{(3,0)}(\stil,\ttil) \\
	&\quad\qquad-\bigg[\frac{\pi^6}{2^5\cdot51975}\Big(\big(\tfrac{11}{36}\stil^{10}\zeta_3+p(\stil,\ttil)\big)\log(-\stil)+(\stil\leftrightarrow\ttil)+(\stil\leftrightarrow\util)\Big)+(\texttt{analytic})\bigg] \\
	&\quad\qquad+\bigg[\frac{\pi^6\zeta_3}{2^{11}\cdot99225}\Big(\stil^9(846\stil^2+\ttil\util)\log^2(-\stil)+(\stil\leftrightarrow\ttil)+(\stil\leftrightarrow\util)\Big)+(\texttt{lower log's})\bigg] \\
	&\quad\qquad-\bigg[\frac{\pi^6\zeta_3}{91125}\Big(\stil^{12}\log^3(-\stil)+\ttil^{12}\log^3(-\ttil)+\util^{12}\log^3(-\util)\Big)+(\texttt{lower log's})\bigg] \\
	&\quad\qquad+\bigg[\frac{\pi^6\zeta_5}{2^{14}\cdot334884375}\Big(\stil^9(1101385\stil^4+1015\stil^2\ttil\util-29\ttil^2\util^2)\stil^{13}\log^2(-\stil)\\
	&\qquad\qquad\qquad\qquad\qquad\qquad\qquad\quad+(\stil\leftrightarrow\ttil)+(\stil\leftrightarrow\util)\Big)+(\texttt{lower log's})\bigg]\bigg) \\
	&\quad+\o(\ap^{17})\,.
\end{split}
\end{align}
Terms at higher orders have leading log's, and have been computed in Section \ref{subsec:genus_3}.

Apart from the rather involved logarithmic structure, note that the sub-leading log at order $\ap^{13}$ receives contributions from three different cuts: $\R^4\vert\partial^6\R^4_2$, $\R^4_1\vert\partial^6\R^4_1$ and $S\vert\partial^{12}\R^4_2$ all contribute at the same order. In particular, the last cut involves the genus-two correction to the $\partial^{12}\R^4$ term whose coefficient is currently not known. With the parametrisation given in \eqref{eq:higher2}, its contribution to the sub-leading log at order $\ap^{13}$ reads
\begin{align}
\begin{split}
	p(\stil,\ttil) &= 64\stil^4\big(2260\stil^6-345\stil^4tu+57\stil^2\ttil^2\util^2-5\ttil^3\util^3\big)D_1\\
	&~~+792\stil^6\big(20\stil^4+5\stil^2tu+2\ttil^2\util^2\big)D_2\,.
\end{split}
\end{align}
Next, we will take a closer look at the origin of the $\stil^{10}\zeta_3$ term, with which the above contribution mixes.

%%%%%%%%%%%%%%%%%%%%%%%%%%%%%%%%%%%%%%%%%%%%%%%%%%%%%%%%%%%%%%%%%%%%%%%%%%%%%%%%%%%%%%%%%%
\subsection{Observation on uniform transcendentality of sub-leading log’s}\label{subsec:transcendentality}
As mentioned above, three different terms contribute to the sub-leading log at order $\ap^{13}$ in \eqref{eq:genus_3_final}. Two of these, namely $\R^4\vert\partial^6\R^4_2$ and $\R^4_1\vert\partial^6\R^4_1$, contribute with a factor of $\zeta_3$. The origin of that zeta-value is $\R^4$ (at genus 0) in the first case, and $\partial^6\R^4_1$ (at genus 1) in the second, while the other two vertices which enter these cuts have transcendental weight 0. Given that $p(\stil,\ttil)$ contributes at the same order, we are tempted to push this logic one step further. Assuming \textit{uniform transcendentality} of the single log at that order, we are led to conjecture the following:\\\vspace{-0.2cm}

\newtheorem{conj}{Conjecture}
\noindent\fbox{\begin{minipage}{1.0\textwidth}
\begin{conj}\label{conj}
	The coefficient of the genus-two correction to the $\partial^{12}\R^4$ term, parametrised by $D_1$ and $D_2$ in equation \eqref{eq:higher2}, are proportional to $\zeta_3$.
\end{conj}
\end{minipage}}\\\vspace{0.1cm}

\noindent We hasten to add that the assumption about uniform transcendentality of sub-leading log's is supported by data from previous orders. For instance, one example involving the unprotected $\partial^{10}\R^4$ term is the sub-leading log at order $\ap^{12}$ in the genus-two expansion \eqref{eq:genus_2_final}. Looking at Table \ref{tab:double_exp}, one sees there are two contributions at that order: $S\vert\partial^{10}\R^4_1$ and $\partial^4\R^4\vert\R^4_1$. And again, both terms happen to come with the same transcendental constant, $\zeta_5$ in this case. In the first term, the $\zeta_5$ is provided by $\partial^{10}\R^4_1$ (at genus 1), in the second term by $\partial^4\R^4$ (at genus 0). As before, we observe that the transcendental weights of these terms at different genera are correlated in precisely such a way that the sub-leading log is of uniform transcendental weight.\footnote{
In contrast, the non-analytic terms at genus one are \textit{by construction} of uniform transcendental weight. This is due to the fact that in the low-energy expansion of the tree-level amplitude, the transcendental weight of the coefficients is directly linked to the order in $\ap$. The same argument implies that also the leading log's at higher genus exhibit uniform transcendentality. On the other hand, as one can see from the coefficients given in eqs. \eqref{eq:higher1}-\eqref{eq:higher3}, the transcendental weight of higher-genus analytic terms appears to be shifted with respect to the assignment at tree-level. For this reason, uniform transcendentality of sub-leading log's is \`a priori not manifest.}
Further instances of the same phenomenon happen at order $\ap^{10}$ at genus three and two (with transcendental weights 3 and 0, respectively).

In all the cases described above, coefficients from seemingly unrelated terms in the low-energy expansion (including unprotected ones!) combined in precisely such a way that certain sub-leading log's at higher genus have the same transcendental weight. Thus, viewed from a different perspective, the observation of uniform transcendentality points towards the idea that the transcendental weight of certain analytic terms is related across different genera and orders in $\ap$. Indeed, from the known analytic terms collected in Section \ref{subsec:known_analytic_terms}, one can discern the following pattern: after absorbing a factor of $\pi^2$ into the genus expansion, terms which are $g_s^2\,\ap^3$ apart share the same transcendentality. Put simply, to find the same transcendental weight one genus higher, the power of $\ap$ needs to be increased by three. Looking at the known analytic terms, we can identify three examples of such ``trajectories'', involving coefficients of increasing transcendental weight:\vspace{-0.1cm}
\begin{itemize}
\item weight 0: $~~\quad\, S~\rightarrow~\,~\R^4_1~\,~\rightarrow~\partial^6\R^4_2~.$\vspace{-0.2cm}
\item weight 3: $~~~\,\R^4~\rightarrow~\partial_6\R^4_1~\rightarrow~\partial^{12}\R^4_2~.$\vspace{-0.2cm}
\item weight 5: $\,\partial^4\R^4~\rightarrow~\partial^{10}\R^4_1~.$\vspace{-0.2cm}
\end{itemize}
In all three cases, the sequence of analytic terms stops because the next term is masked by a non-analytic contribution, which prevents us from accessing the underlying analytic term. Viewed in this way, Conjecture \ref{conj} is simply the continuation of the weight 3 trajectory to the third term $\partial^{12}\R^4_2$. 

To conclude, it would be fascinating to understand whether all these observations are just an accident at low orders in $\ap$, or whether they extend to more terms and higher orders. The fact that the patterns we observed extend beyond just the protected higher-genus corrections is encouraging. A natural question to then ask is if imposing uniform transcendentality of sub-leading log's in a more systematic manner can teach us anything new about the unprotected higher-genus analytic terms and their modular properties. We hope that these questions will be addressed in the near future.

%%%%%%%%%%%%%%%%%%%%%%%%%%%%%%%%%%%%%%%%%%%%%%%%%%%%%%%%%%%%%%%%%%%%%%%%%%%%%%%%%%%%%%%%%%
\section{Application II: the type I open string amplitude}\label{sec:open_strings}
%%%%%%%%%%%%%%%%%%%%%%%%%%%%%%%%%%%%%%%%%%%%%%%%%%%%%%%%%%%%%%%%%%%%%%%%%%%%%%%%%%%%%%%%%%
We will now consider the four-point amplitude in type I string theory. The logic here is very similar to the closed string case, except for the additional complication that we need to deal with the non-trivial colour structure supported by the open string amplitude. As we will see in a moment, perhaps the cleanest way to take the gauge group dependence into account is to decompose the amplitude by introducing orthogonal projectors onto invariant subspaces.
Much like the partial waves themselves, the underlying idea is to move to an orthogonal basis which diagonalises the sum over intermediate states. With this in mind, our strategy is as follows: 1) project the amplitude onto irreps of the gauge group; 2) compute the partial-wave coefficients in each channel separately; 3) glue them to obtain the iterated $s$-channel cut. \\\vspace{-0.2cm}

We begin by recalling some known features of the four-point open string amplitude in type I string theory.
The loop expansion in terms of the open string coupling $g_s$  reads\footnote{Here we will use the terminology loop expansion, since in the case of open unoriented strings the genus of the worldsheet does not necessarily correspond to the order in $g_s$. Instead, the latter is counted by the Euler characteristic, which is specified by three parameters, see the discussion in subsection \ref{subsec:stringtop} for more details.}
\begin{align}\label{eq:genus_exp_open}
	\mathcal{V}^{I_1 I_2 I_3 I_4}(k_i,\zeta_i) = \frac{g_{\text{YM}}^2}{\alpha'^3} \times \alpha'^2 \mathcal{F}^4 \, \sum_{h=0}g_s^{2h+2} V^{(h);I_1 I_2 I_3 I_4}(s, t;\alpha')\,,
\end{align}
where the relation between $g_{\text{YM}}$ and the gravitational coupling in type I string theory is given by \cite{Polchinski:1998rr}
\begin{equation}
g_{\text{YM}}^2= \sqrt{2}(2\pi)^{\frac{7}{2}}\ap\kappa_{10} = (2\pi)^7 \alpha'^3 \,.
\end{equation}
As in the four-point closed amplitude of the IIB theory, the polarisation dependence is entirely captured by the invariant $\mathcal{F}^4$, which is the supersymmetric completion   \cite{Berkovits:2004px} of the well-known bosonic tensor $t_8 = (F_{\mu\nu}F^{\nu\rho}F_{\rho\sigma}F^{\sigma\mu})-\frac{1}{4}(F_{\mu\nu}F^{\mu\nu})^2\,$ \cite{Green:1982sw}. It satisfies the self-replicating property \cite{Eberhardt:2022zay}:
\begin{align}\label{eq:YMsewing_relation}
	\sum_{\rm 2\, pt \,states}\mathcal{F}^4_{1,2\rightarrow {\rm 2\, pt }}\,\mathcal{F}^4_{ {\rm 2\, pt } \rightarrow 3,4} = \frac{s^2}{2}\,\mathcal{F}^4_{1,2\rightarrow 3,4}\,.
\end{align}

Concerning the dependence of the $h$-loop amplitude $V^{(h);I_1 I_2 I_3 I_4}$ on the colour indices $I_i$, the standard decomposition utilises the so-called trace-basis. For four-point amplitudes (at any loop order), this basis consists of single- and double-trace terms only. The amplitude can thus be written as
\begin{align}\label{eq:trace_basis}
\begin{split}
V^{(h);I_1 I_2 I_3 I_4} = \, &   V_{1234}^{(h)}\,\Tr_{1234}  +(\texttt{crossing}) \,  + \\
		& V_{12;34}^{(h)}\,\Tr_{12;34} +(\texttt{crossing})\,,\\
\end{split}
\end{align}
where $\Tr_{1234}\equiv\Tr[T^{I_1} T^{I_2}T^{I_3}T^{I_4}]$, $\Tr_{12;34}\equiv\Tr[T^{I_1} T^{I_2}]\Tr[T^{I_3}T^{I_4}]=\delta^{I_1 I_2} \delta^{I_3 I_4}$. Here $T^{I_i}$ are the generators of $SO(N)$ and $(\texttt{crossing)}$ stands for two more inequivalent orientations of the single- and double-traces. Note that in type I string theory absence of anomalies requires the gauge group to be $SO(32)$, but it is useful to keep the rank generic. As we are going to see later, this will help to disentangle the various topologies contributing to the open string scattering in the final results.

%%%%%%%%%%%%%%%%%%%%%%%%%%%%%%%%%%%%%%%%%%%%%%%%%%%%%%%%%%%%%%%%%%%%%%%%%%%%%%%%%%%%%%%%%%
\subsection{Decomposition into orthogonal projectors}
The trace-basis mentioned above has the drawback that it is not an orthogonal basis. Thus the sum over intermediate states requires some care as one cannot just naively multiply tree-level partial-wave coefficients associated to a given colour-ordered amplitude.
A more suitable way to take the colour structure into account is to perform a change of a basis and work instead in an orthogonal basis, where the multiplication of partial-wave coefficients can instead be performed independently in each invariant subspace.
This is provided by the decomposition of the amplitude into projectors $\mathbb{P}_{\bf{a}}$, which project the amplitude onto certain irreps of the gauge group.
We will be dealing with external adjoint fields, thus we need to consider the irreps in the tensor product $\texttt{adj}\otimes \texttt{adj}$:
\begin{equation}
\mathcal{V}^{I_1 I_2 I_3 I_4} =\sum_{\bf{a}\in \texttt{adj}\otimes \texttt{adj}} \mathcal{V}_{\bf{a}} \, \mathbb{P}_{\bf{a}}^{I_1 I_2 I_3 I_4}.
\end{equation} 
The bold-font index ${\bf{a}}$ runs over the 6 irreps in the tensor product\footnote{Note that this decomposition is valid only for $SO(N)$ with $N\geq4$ and $N\neq8$. In those other cases, the number of irreps differs from 6.}
\begin{align}\label{tensordecomp}
& \texttt{adj}\otimes \texttt{adj} =   {\bf{1}}   \oplus \texttt{sym}_1 \oplus \texttt{sym}_2\oplus \texttt{sym}_3 \oplus   \texttt{adj}    \oplus   \texttt{asym}\,,
\end{align}
with their dimensions being
\begin{align}
\begin{split}
\text{dim}[\texttt{sym}_1] &=  \tfrac{(N{-}1)(N{+}2)}{2}\,,\\ 
\text{dim}[\texttt{sym}_2] &= \tfrac{(N{-}3)(N{-}2)(N{-}1)N}{24}\,,\\
\text{dim}[\texttt{sym}_3] &=  \tfrac{(N{-}3)N(N{+}1)(N{+}2)}{12}\,,\\ 
\text{dim}[\texttt{adj}]   &= {\tfrac{N(N{-}1)}{2}}\,,\\
\text{dim}[\texttt{asym}]  &= \tfrac{N(N{-}1)(N{-}3)(N{+}2)}{8}\,.
\end{split}
\end{align}
Note that the projectors satisfy the two properties
\begin{align}
	\mathbb{P}_\mathbf{a}^{I_1I_2I_3I_4}\mathbb{P}_\mathbf{b}^{I_4I_3I_5I_6}=\delta_{\mathbf{a} , \mathbf{b}}\,\mathbb{P}_\mathbf{b}^{I_1I_2I_5I_6}\,,\qquad \Tr(\mathbb{P}_\mathbf{a})=\mathbb{P}_\mathbf{a}^{I_1I_2I_2I_1}=\text{dim}(\mathbf{a})\,.
\end{align} 
In particular, the idempotency condition allows us to consider contributions from the various irreps involved in the cut separately. This is in contrast with the trace-basis, where the product of traces generically gives rise to a sum of both single- and double-trace contributions. 

Recall from \eqref{eq:trace_basis} that a four-point amplitude can be written as a sum of single- and double-trace terms.
Thus, all we need to know is the decomposition of single- and double-trace structures into the above basis of irreps. This is readily computed and reads
\begin{align}\label{tracetoproj}
\begin{split}
\Tr_{1234} &= \big(\tfrac{N-1}{2},\tfrac{N-2}{4},0,0,\tfrac{N-2}{4},0 \big)\,,\\ 
\Tr_{1243} &= \big(\tfrac{N-1}{2},\tfrac{N-2}{4},0,0,\tfrac{2-N}{4},0  \big)\,,\\ 
\Tr_{1324} &= \big(\tfrac{1}{2},\tfrac{1}{2},-1,\tfrac{1}{2},0,0 \big)\,,\\ 
\Tr_{12;34} &= \big(\tfrac{N(N-1)}{2},0,0,0,0,0  \big)\,,\\ 
\Tr_{13;24} &= \big(1,1,1,1,-1,-1 \big)\,,\\ 
\Tr_{14;23} &=\big( 1,1,1,1,1,1 \big)\,,
\end{split}
\end{align}
where the vectors are ordered as in the decomposition \eqref{tensordecomp}. The first four entries are thus the symmetric irreps, and the last two are antisymmetric ones.

%%%%%%%%%%%%%%%%%%%%%%%%%%%%%%%%%%%%%%%%%%%%%%%%%%%%%%%%%%%%%%%%%%%%%%%%%%%%%%%%%%%%%%%%%%
\subsection{Digression: worldsheet topologies}\label{subsec:stringtop}
Before diving into the computation of the leading log's, we would like to make a little detour on the rank dependence of the open string amplitudes.
There is one important advantage of decomposing the amplitude into projectors, which is perhaps worth stressing: this formalism automatically takes care of \emph{all} contributions coming from different worldsheet topologies and from the different ways of placing the open strings on the boundaries at each loop order. This is in contrast with direct worldsheet computations \cite{Eberhardt:2022zay}, where each topology needs to be treated separately. This adds additional complications even at one-loop order. Similarly, the methods of \cite{Edison:2021ebi} might not be straightforward to generalise beyond planar oriented topologies.\\\vspace{-0.2cm}

Recall that the inequivalent topologies are labelled by triplets $(g,b,c)$, where $g$ is the number of handles, $b$ the number of boundaries, and $c$ the number of cross-caps, with the Euler characteristic computed by $\chi=2-2g-b-c$.\footnote{This classification is in fact redundant. In order to obtain each topology exactly once, one can let only one of $g$ and $c$ to be non-zero, see the discussion in e.g. \cite{Polchinski:1998rq}.} In terms of the loop order $h$, we have $h=1-\chi$.

\begin{figure}[t]
\begin{tikzpicture}[scale=1.5]
\newcommand \y {4};
% left
\draw[very thick, gray] (0,0) circle (1);
\draw[very thick, gray]  (0,0) circle (0.5);
\draw (0.43,0.9) node[cross] {};
\draw (-0.43,0.9) node[cross] {};
\draw (0.43,-0.9) node[cross] {};
\draw (-0.43,-0.9) node[cross] {};
\node at (-0.6, 1.0) {1};
\node at (0.6, 1.0) {2};
\node at (-0.6, -1.0) {4};
\node at (0.6, -1.0) {3};
% middle
\draw[very thick, gray]  (4,0) circle (1);
\draw[very thick, gray]  (4,0) circle (0.5);
\draw (0.43+\y , 0.9) node[cross] {};
\draw (-0.43+\y , 0.9) node[cross] {};
\draw (0.3+\y , - 0.40) node[cross] {};
\draw (-0.3+\y , -0.40) node[cross] {};
\node at (-0.6+\y, 1.0) {1};
\node at (0.6+\y, 1.0) {2};
\node at (-0.5+\y, -0.6) {4};
\node at (0.5+\y, -0.6) {3};
% right
\draw[very thick , gray] (2*\y , 0) circle (1);
\draw[very thick, gray] (2*\y , 0.2) circle (.5);
\fill[white] (2*\y-0.35,0) rectangle (2*\y+0.3,1.3);
\draw[very thick, gray, out=30, in=135, looseness=.8] (-.425+2*\y, .906) to (.33+2*\y,.575);
\draw[very thick, gray, out=165, in=48, looseness=.8] (.32+2*\y, .948) to (-.38+2*\y,.525);
 \draw (0.43+2*\y,0.9) node[cross] {};
\draw (-0.43+2*\y,0.9) node[cross] {};
\draw (0.43+2*\y,-0.9) node[cross] {};
\draw (-0.43+2*\y,-0.9) node[cross] {};
\node at (-0.6+2*\y, 1.0) {1};
\node at (0.6+2*\y, 1.0) {2};
\node at (-0.6+2*\y, -1.0) {4};
\node at (0.6+2*\y, -1.0) {3};
\end{tikzpicture}
\caption{One-loop topologies contributing to open string scattering. Thick lines represent boundaries. The planar cylinder (left) contributes to single-trace terms of order $ N$, while double-trace terms emerge from the non-planar cylinder (middle). Finally, the Moebius topology (right) contributes to single-trace terms of order $N^0$.}
\label{figure:onelooptopologies}
\end{figure}
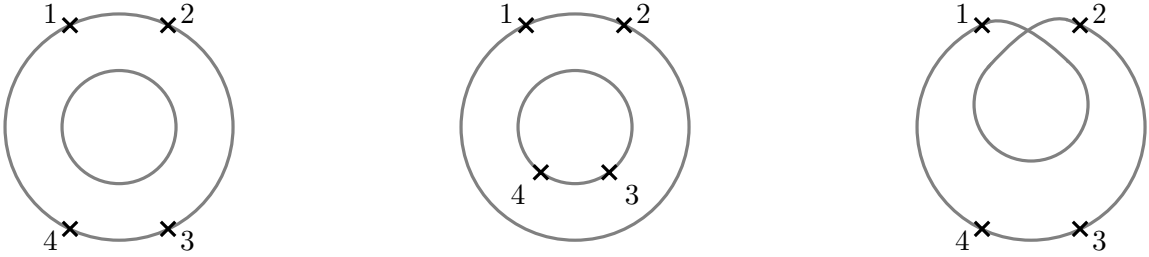

For concreteness, in the following we discuss the $N$-counting for the $\log^h(-\tilde{s})$ coefficient in the $1234$ ordering, namely $f^{(h,n)}_{1234}$ and $f^{(h,n)}_{12;34}$, the other cases being related by crossing.
Beginning at one loop ($\chi=0$), there are two inequivalent topologies with boundaries, namely the cylinder $(0,2,0)$, and the Moebius strip  $(0,1,1)$, as shown in Figure \ref{figure:onelooptopologies}.
Single-trace terms of order $N$ come from placing all open strings on one boundary of the cylinder; double-trace terms also arise from the cylinder but this time placing two strings on one boundary and two strings on the other boundary. Lastly, single trace terms of order $N^0$ arise from the Moebius topology.
In Section \ref{sec:leading_logs_open}, we will see that the one-loop amplitude has indeed precisely this structure, c.f.	\eqref{eq:singletraceonel}-\eqref{eq:doubleoneloop}.   

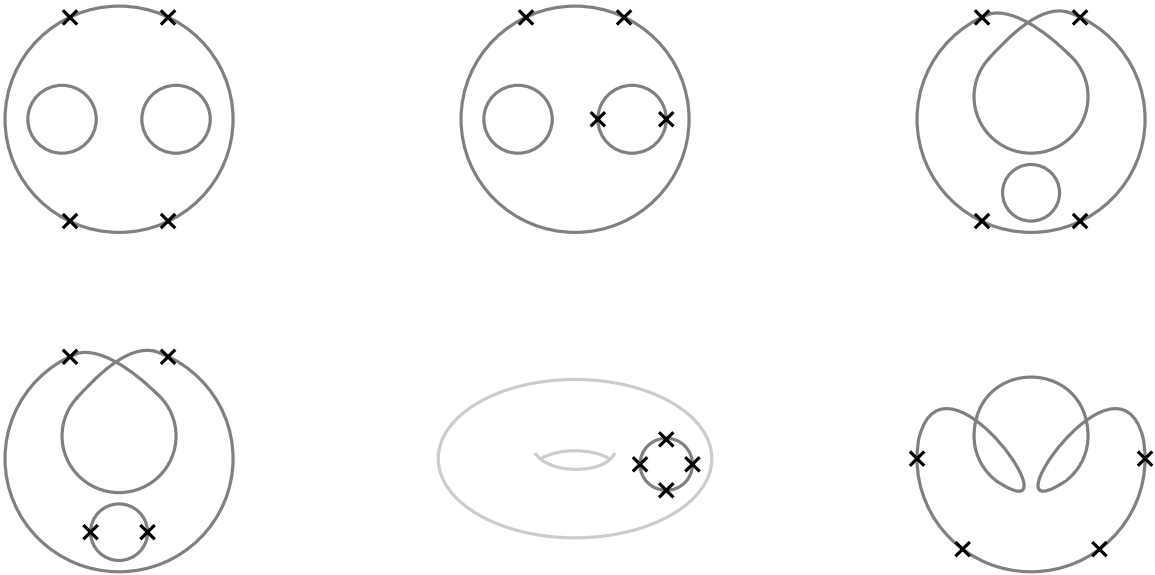
\begin{figure}[t]
\begin{tikzpicture}[scale=1.5]
\draw[very thick, gray]  (0,0) circle (1);
\draw[very thick, gray]   (-0.5,0) circle (0.3);
\draw[very thick, gray]   (0.5,0) circle (0.3);
\newcommand \y {4};
\newcommand \z {3};

 \draw (0.43,0.9) node[cross] {};
  \draw (-0.43,0.9) node[cross] {};
   \draw (0.43,-0.9) node[cross] {};
  \draw (-0.43,-0.9) node[cross] {};

\draw[very thick, gray]  (\y,0) circle (1);
\draw[very thick, gray]   (-0.5+\y,0) circle (0.3);
\draw[very thick, gray]   (0.5+\y,0) circle (0.3);

 \draw (0.43+\y , 0.9) node[cross] {};
  \draw (-0.43+\y , 0.9) node[cross] {};
  \draw (0.2+\y , 0) node[cross] {};
    \draw (0.8+\y , 0) node[cross] {};
  
 \draw[very thick ,gray] (2*\y , 0) circle (1);
\draw[very thick, gray] (2*\y , 0.2) circle (.5);
          \fill[white] (2*\y-0.35,0) rectangle (2*\y+0.3,1.3);
          
 \draw[very thick, gray, out=30, in=135, looseness=.8] (-.425+2*\y, .906) to (.33+2*\y,.575);
 
 \draw[very thick, gray, out=165, in=48, looseness=.8] (.32+2*\y, .948) to (-.38+2*\y,.525);
 
 \draw (0.43+2*\y,0.9) node[cross] {};
  \draw (-0.43+2*\y,0.9) node[cross] {};
   \draw (0.43+2*\y,-0.9) node[cross] {};
  \draw (-0.43+2*\y,-0.9) node[cross] {};

 \draw[very thick ,gray] (2*\y , -.65) circle (.25);
 
 \draw[very thick ,gray] (0 , -\z) circle (1);
\draw[very thick, gray] (0, 0.2-\z) circle (.5);
          \fill[white] (0-0.35,0-\z) rectangle (+0.3,1.3-\z);
          
  \draw[very thick, gray, out=30, in=135, looseness=.8] (-.425, .906-\z) to (.33,.575-\z);
  
  \draw[very thick, gray, out=165, in=48, looseness=.8] (.32, .948-\z) to (-.38,.525-\z);
 
 \draw (0.43,0.9-\z)  node[cross] {};
  \draw (-0.43,0.9-\z)  node[cross] {};
 \draw[very thick ,gray] (0 , -.65-\z) circle (.25);
 \draw (0.25,-0.65-\z)  node[cross] {};
  \draw (-0.25,-0.65-\z)  node[cross] {};
 
 \draw[very thick,gray!40] (\y , -\z) ellipse (1.2 and 0.7);
  \draw[very thick, ,gray!40, out=30, in=150, looseness=.8] (\y-0.3, -\z) to  (\y+0.3, -\z);
\draw[very thick, ,gray!40,  out=300, in=245, looseness=.8] (\y-0.35, -\z+.05) to  (\y+0.35, -\z+.05);
 \draw[very thick ,gray] (\y+.8, -.05-\z) circle (.23);
 \draw (\y+.8, 0.17-\z)  node[cross] {};
  \draw (\y+.8,-0.28-\z)  node[cross] {};
  \draw (\y+.57, -.05-\z)  node[cross] {};
  \draw (\y+1.03, -.05-\z)  node[cross] {};
  
     \draw[very thick ,gray] (2*\y, -\z) circle (1);
\draw[very thick, gray] (2*\y, 0.2-\z) circle (.5);
  \fill[white] (-1.1+2*\y,0.2-\z)  rectangle (1.1+2*\y,1.3-\z);
 \fill[white] (-0.3+2*\y,-0.5-\z)  rectangle (.3+2*\y,0.3-\z);

 \draw[very thick, gray, out=75, in=-34, looseness=3] (-0.98+2*\y,0.194-\z) to (-0.308+2*\y,-.1945-\z);
 \draw[very thick, gray, out=105, in=214, looseness=3] (0.98+2*\y,0.194-\z) to (0.308+2*\y,-.1945-\z);

 \draw[very thick, gray, out=90, in=90, looseness=1.8] (-0.5+2*\y,0.1945-\z) to (0.5+2*\y,0.1945-\z);

 \draw (2*\y - 1, -\z)  node[cross] {};
  \draw (2*\y + 1, -\z)  node[cross] {};
  \draw (2*\y -0.6, -\z-0.8)  node[cross] {};
  \draw (2*\y +0.6, -\z-0.8)  node[cross] {};
\end{tikzpicture}
\caption{Two-loop topologies contributing to open string scattering. We assume clockwise ordering of the insertions as in the one-loop case. From top left to bottom right: $(0,3,0)$ topology contributing to single-trace terms of order $N^2$; $(0,3,0)$ topology contributing to double-trace terms of order $N$; $(0,2,1)$ topology contributing to single-trace terms of order $N$; $(0,2,1)$ topology contributing to double-trace terms of order $N^0$;  $(1,1,0)$ topology contributing to single-trace terms of order $N^0$; $(0,1,2)$ topology contributing to single-trace terms of order $N^0$.}
\label{figure:twolooptopol}
\end{figure}

Analogously, at two loops ($\chi=-1$) the single-trace terms are quadratic in $N$, and one double-trace term is linear. These are in correspondence with two-loop topologies, see Figure \ref{figure:twolooptopol}. For example, the quadratic term in $N$ arises from the topology $(0,3,0)$ with all strings on one boundary, the linear term in $N$ in the single trace arises from a Moebius strip with an additional boundary $(0,2,1)$, and so on.
Note that the order $N^0$ term in the single-trace term arises from two contributions: a double cross-cap $(0,1,2)$ and a torus with a boundary $(1,1,0)$.

%%%%%%%%%%%%%%%%%%%%%%%%%%%%%%%%%%%%%%%%%%%%%%%%%%%%%%%%%%%%%%%%%%%%%%%%%%%%%%%%%%%%%%%%%%
%%%%%%%%%%%%%%%%%%%%%%%%%%%%%%%%%%%%%%%%%%%%%%%%%%%%%%%%%%%%%%%%%%%%%%%%%%%%%%%%%%%%%%%%%%

%%%%%%%%%%%%%%%%%%%%%%%%%%%%%%%%%%%%%%%%%%%%%%%%%%%%%%%%%%%%%%%%%%%%%%%%%%%%%%%%%%%%%%%%%%
%%%%%%%%%%%%%%%%%%%%%%%%%%%%%%%%%%%%%%%%%%%%%%%%%%%%%%%%%%%%%%%%%%%%%%%%%%%%%%%%%%%%%%%%%%

%%%%%%%%%%%%%%%%%%%%%%%%%%%%%%%%%%%%%%%%%%%%%%%%%%%%%%%%%%%%%%%%%%%%%%%%%%%%%%%%%%%%%%%%%%
\subsection{Partial-wave expansion of the open string tree-level amplitude}
The computation of leading log's via unitarity cuts relies only on the tree-level amplitude, which in the case of type I string theory is given by the well-known Veneziano formula. Note that the colour-structure of the tree-level amplitude is special in that the double-trace terms vanish. In the notation of \eqref{eq:trace_basis}, we thus have
\begin{align}\label{Venezianoflat}
\begin{split}
V^{(0)}_{1234}(s,t)  &= - 8  \alpha'^3\, \frac{\Gamma(-\tilde{s})\Gamma(-\tilde{t})}{\Gamma(1+\tilde{u})} = 8\alpha'^3 \left[- \frac{1}{\tilde{s}\tilde{t}}+\zeta_2 - \zeta_3 \tilde{u} + \frac{\zeta_4}{4}(4\util^2-7\stil\ttil) +\ldots  \right],\\
V^{(0)}_{12;34}(s,t) &= 0\,,
\end{split}
\end{align}
and note that $V^{(0)}_{1234}(s,t)$ is symmetric under $s\leftrightarrow t$ exchange.
The other two orientations of the single-trace term are given in terms of $V^{(0)}_{1234}(s,t)$ by
\begin{align}
V^{(0)}_{1243} (s,t) =V^{(0)}_{1234} (s,u)\,,\quad V^{(0)}_{1324} (s,t) = V^{(0)}_{1234} (t,u)\, . 
\end{align}
As advocated before, our first task is to change from the trace-basis to the basis of irreps. By means of \eqref{tracetoproj}, one has\footnote{The horizontal line is just a visual aid to separate the symmetric from the antisymmetric irreps.}
\begin{equation}
V_{\bf{a}}^{(0)} = \frac{1}{2} 
\begin{pmatrix}
 (N-1) (V^{(0)}_{1234}+V^{(0)}_{1243})+V^{(0)}_{1324} \\
\frac{N-2}{2} (V^{(0)}_{1234}+V^{(0)}_{1243})+V^{(0)}_{1324}  \\
- 2 V^{(0)}_{1324}  \\
V^{(0)}_{1324} \\
 \hline
 \frac{N-2}{2}( V^{(0)}_{1234}-V^{(0)}_{1243} )  \\
0  \\
\end{pmatrix},
\end{equation}
where the symmetric (antisymmetric) irreps are symmetric (antisymmetric) under $t\leftrightarrow u$ exchange, as dictated by crossing symmetry.
Having switched to an orthogonal basis, we can now perform the partial-wave decomposition in each channel separately. Expanding furthermore in small $\ap$, we have
\begin{equation}
V_{\bf{a}}^{(0)}\vert_{\ap^{n+1}} = \frac{1}{s^3}\sum_{\ell} \epsilon_{\bf{a},\ell}^{(n)}\, P_{\ell}(z)\,,
\end{equation}
which defines the tree-level partial-wave coefficients $\epsilon_{\bf{a},\ell}^{(n)}$. Note that for symmetric (antisymmetric) irreps $\bf{a}$, the sum over spin $\ell$ runs over even (odd) integers.

For the first term in the low-energy expansion, given by the field-theory term $F^2$, the partial-wave coefficients take the form
\begin{equation}\label{eq:treepartialYM}
F^2:\qquad\epsilon_{\bf{a},\ell}^{(0)} = \frac{\tilde{s}}{2^{5} \pi^4{(\ell+1)_{6}}}
\begin{pmatrix}
N-2  \\
\frac{N-4}{2}  \\
2  \\
-1\\\hline
\frac{N-2}{2}  \\
0  \\
\end{pmatrix}.
\end{equation}
As expected, the $\epsilon_{\bf{a},\ell}^{(0)}$ have infinite spin support. Note also that its $\ell$-dependence is the same as the closed string case; this is simply a consequence of the fact that supergravity and field theory amplitudes only differ by a power of $\tilde{s}$ and the latter does not affect the spin dependence of the partial-wave coefficients.

On the other hand, higher-derivative corrections are polynomials in the Mandelstam variables, and thus the corresponding partial-wave coefficients $\epsilon^{(n\geq2)}_{\bf{a},\ell}$ are spin-truncated.
For the first two string corrections for example, given by the $F^4$ and $\partial^2F^4$ higher-derivative corrections, the partial-wave coefficients read
\begin{equation}\label{eq:epsilonopen2}
F^4:\qquad\epsilon_{{\bf{a}},\ell}^{(2)} = \frac{\tilde{s}^3\,\zeta_2}{2^{10} \cdot 105  \pi^4}\delta_{\ell,0}
\begin{pmatrix}
2N-1 \\
N-1 \\
-2\\
1\\\hline
0  \\
0  \\
\end{pmatrix},
\end{equation}
and
\begin{equation}\quad 
\partial^2F^4:\qquad\epsilon_{{\bf{a}},\ell}^{(3)} = \frac{\tilde{s}^4\,\zeta_3}{2^{10} \cdot 105 \pi^4}
\begin{pmatrix}
(N-2)\delta_{\ell,0} \\
\frac{1}{2}(N-4)\delta_{\ell,0} \\
2 \delta_{\ell,0}\\
-\delta_{\ell,0} \\\hline
\frac{1}{18}( N-2) \delta_{\ell,1} \\
0  \\
\end{pmatrix}.
\end{equation}
Higher-order string corrections are easily obtained by means of \eqref{eq:pw_monomial}.

%%%%%%%%%%%%%%%%%%%%%%%%%%%%%%%%%%%%%%%%%%%%%%%%%%%%%%%%%%%%%%%%%%%%%%%%%%%%%%%%%%%%%%%%%%
\subsection{Explicit results for leading log's}\label{sec:leading_logs_open}
In the rest of this section, we will compute the iterated $s$-channel cut by taking products of the tree-level partial-wave coefficients $\epsilon_{{\bf{a}},\ell}^{(n)}$ in each invariant subspace. Following a similar notation as in the closed string case, we wish to compute the leading-log coefficient,
\begin{align}
	f^{(h)}_{{\bf{a}}}(s,t;\ap)\equiv V^{(h)}_{{\bf{a}}}(s,t;\ap)\vert_{\log^h(-\stil)}\,,
\end{align}
in a low-energy expansion:
\begin{equation}
 f_{{\bf{a}}}^{(h)}(s,t;\ap) = \ap^3 f_{{\bf{a}}}^{(h)} (\tilde{s},\tilde{t}) = \ap^3\sum_{n\geq 0 } f_{{\bf{a}}}^{(h,n)}(\tilde{s},\tilde{t})=\sum_{n\geq 0 } (\ap)^{3h+n+1} f_{{\bf{a}}}^{(h,n)}(s,t)\,,
\end{equation}
with the  $f_{{\bf{a}}}^{(h,n>0)}(\tilde{s},\tilde{t})$ being homogeneous polynomials of degree $3h+n-2$. Here, the superscript $(h,n)$ denotes the loop-order $h$ and $\ap$-order $3h+n+1$. It is important to note that this is different from the $\ap$-counting in the closed string case. Next, we employ the partial-wave expansion on each $f_{{\bf{a}}}^{(h,n)}(\tilde{s},\tilde{t})$, order by order in $\ap$: 
\begin{equation}\label{eq:fopenstrings}
\alpha'^3 f_{{\bf{a}}}^{(h,n)}(\tilde{s},\tilde{t})=\frac{1}{s^3}\sum^\infty_{\ell}\sum^\infty_{ n=0}\epsilon_{\bf{a},\ell}^{(h,n)} \,P_\ell(z)\,,
\end{equation} 
where $\epsilon_\ell^{(0,n)}\equiv \epsilon_\ell^{(n)}$. Adapting eq. \eqref{ddiscLloops2} and matching the $\ap$-expansions on both sides we find that $\epsilon_{\bf{a},\ell}^{(h,n)}$ is fully fixed by tree-level data:
\begin{empheq}[box=\fbox]{equation}
\begin{aligned}\label{eq:epsilonopen}
\epsilon_{\bf{a},\ell}^{(h,n)}= \left(-\frac{1}{4 \pi} \times \frac{(2\pi)^7\tilde{s}^2}{2}\right)^h \sum_{\sigma_{n, h}}\epsilon_{\bf{a},\ell}^{(n_1)}\epsilon_{\bf{a},\ell}^{(n_2)}\cdots \epsilon_{\bf{a},\ell}^{(n_{h{+}1})}\,,
\end{aligned}
\end{empheq}
where we recall that $\sigma_{n,h}$ represents all possible solutions to $\sum_{i=1}^{h+1} n_i=n$,
and the extra factor of $\frac{1}{2} (2\pi)^7\tilde{s}^2$ arises from the kinematical prefactor in \eqref{eq:genus_exp_open} together with the iteration of the self-replicating property \eqref{eq:YMsewing_relation}. The structure of leading log's implied by this gluing operation is summarised in Table \ref{tab:open}. 

%%%%%%%%%%%%%%%%%%%%%%%%%%%%%%%%%%%%%%%%%%%%%%%%%%%%%%%%%%%%%%%%%%%%%%%%%%%%%%%%%%%%%%%%%%
\begin{table}
\begin{center}
\renewcommand\leading[1]{#1}
\def\cwidthmax{2.0cm}
\def\cwidth{1.7cm}
\def\cwidthalt{1.4cm}
\def\cwidthaltt{0.9cm}
\def\cwidthalttt{0.7cm}
$\begin{array}{|c|>{$}C{\cwidthalttt}<{$}|>{$}C{\cwidthalttt}<{$}|>{$}C{\cwidthalttt}<{$}|>{$}C{\cwidthaltt}<{$}|>{$}C{\cwidthaltt}<{$}|>{$}C{\cwidthaltt}<{$}|>{$}C{\cwidthalt}<{$}|>{$}C{\cwidthalt}<{$}|>{$}C{\cwidthalt}<{$}|>{$}C{\cwidth}<{$}|>{$}C{\cwidthalttt}<{$}}\hline
{(\ap)^n} & 1 & 2 & 3 & 4 & 5 & 6 & 7 & 8 & 9 & 10 &\cdots \\ \hline
& & & & & & & & & & &\\[-11pt]
\text{\,tree-level\,} & \FS &  & F & _2F & _4F & _6F & _8F & _{10}F & _{12}F & _{14}F & \cdots  \\[2pt] \hline
& & & & & & & & & & &\\[-11pt]
\text{1-loop} &  &  & \times & \leading{\FS\vert\FS} & \times & \leading{\FS\vert F} & \leading{\FS\vert_2F} & \leading{\begin{array}{@{}c@{}}\FS\vert_4F \\ F\vert F \end{array}} & \leading{\begin{array}{@{}c@{}}\FS\vert_6F \\ F\vert_2F \end{array}} & \leading{\begin{array}{@{}c@{}}\FS\vert_8F \\ F\vert_4F \\ _2F\vert_2F \end{array}} & \cdots \\[17pt] \hline
& & & & & & & & & & &\\[-11pt]
\text{2-loop} &  &  & \times & \times & \times & \times & \leading{\FS\vert\FS\vert\FS} & \times & \leading{\FS\vert\FS\vert F} & \leading{\FS\vert\FS\vert_2F}  & \cdots \\[2pt] \hline
& & & & & & & & & & &\\[-11pt]
\text{3-loop} &  &  & \times & \times & \times & \times & \times & \times & \times & \leading{\FS\vert\FS\vert\FS\vert\FS} & \cdots \\[2pt] \hline
\vdots &  &  & \vdots & \vdots & \vdots & \vdots & \vdots & \vdots & \vdots & \vdots & \ddots
\end{array}$
\end{center}\vspace{-0.5cm}
\caption{Leading logarithmic contributions to the open string amplitude. Here $\FS\equiv F^2$ stands for the tree-level field theory vertex, whereas $_aF\equiv\partial^aF^4$ denotes insertions of higher-derivative contact vertices. A cross indicates that there is no leading log at that order in $\ap$, but sub-leading log's or analytic terms may be present. To the best of our knowledge, for type I string theory no non-renormalisation theorems are known.%
}\label{tab:open}
\end{table}
%%%%%%%%%%%%%%%%%%%%%%%%%%%%%%%%%%%%%%%%%%%%%%%%%%%%%%%%%%%%%%%%%%%%%%%%%%%%%%%%%%%%%%%%%%

With the $s$-channel leading log's at hand, we can then use crossing symmetry to also determine the leading log's in the other orientations. In the basis of irreps, this is done by using the crossing matrices $F_t$ and $F_u$. Their explicit expressions for $SO(N)$ are recorded in Appendix \ref{sec:crossingmatrices}. We then arrive at the following logarithmic structure of the open string amplitude in the projector basis:
\begin{align}\label{eq:veneziano_full}
	V_{\bf{a}}^{(h)} &= \ap^3 V_{\bf{a}}^{(h,0)} \notag\\
	&~+\ap^3\left(\sum_{n=2}^\infty f_{\bf{a}}^{(h,n)}(\stil,\ttil)\log^{h}(-\stil)+  (F_u)_{{\bf{a}}}^{\,\bf{b}}  f_{\bf{b}}^{(h,n)}(\ttil,\stil)\log^h(-\ttil)+ (F_t)_{{\bf{a}}}^{\,\bf{b}}  f_{\bf{b}}^{(h,n)}(\util,\ttil)\log^h(-\util)\right)\notag\\[3pt]
	&~+ (\texttt{lower log's}) + (\texttt{analytic})\,,
\end{align}
where \texttt{lower log's} again stands schematically for all terms with no leading log contribution in any orientation. From there, one can easily change back to the more familiar trace-basis by inverting the relations \eqref{tracetoproj}.

%%%%%%%%%%%%%%%%%%%%%%%%%%%%%%%%%%%%%%%%%%%%%%%%%%%%%%%%%%%%%%%%%%%%%%%%%%%%%%%%%%%%%%%%%%
\subsubsection{One loop}
Let us now apply the above machinery to the one-loop amplitude $V^{(1)}_{{\bf{a}}}$. 
The field-theory one-loop discontinuity $f_{\bf{a}}^{(1,0)}$ is obtained by squaring \eqref{eq:treepartialYM} and performing the (infinite) sum over partial waves \eqref{eq:fopenstrings}. Compared to the supergravity case, we now need to consider the sum over both even and odd spins. One has
\begin{align}
\begin{split}
\frac{1}{\stil^3}\sum_{\ell \,\,\text{even}} \Big(\frac{\stil}{(\ell+1)_6}\Big)^2 P_{\ell}(z) &= - \frac{8 \pi^4}{15 } \frac{1}{\tilde{s}^2} \big( \mathcal{B}(\stil,\ttil) +\mathcal{B}(\stil,\util)  \big)\,, \\ 
\frac{1}{\stil^3}\sum_{\ell \,\,\text{odd}} \Big(\frac{\stil}{(\ell+1)_6}\Big)^2  P_{\ell}(z)  &= - \frac{8 \pi^4}{15} \frac{1}{\tilde{s}^2} \big( \mathcal{B}(\stil,\ttil) -\mathcal{B}(\stil,\util)  \big)\,,
\end{split}
\end{align}  
where we recall for convenience the $s$-channel discontinuity of the 10d one-loop box integral
\begin{equation}
\mathcal{B}(\stil,\ttil) =  \frac{\stil^2}{\util^3} \Big(2\ttil^2 \log(-\stil/\ttil)  -\util(\stil+3\ttil)\Big).
\end{equation}
By taking into account the additional factors originating from \eqref{eq:treepartialYM} and \eqref{eq:epsilonopen}, we find:
\begin{equation}
F^2\vert F^2:\qquad f_{\bf{a}}^{(1,0)} = \frac{\pi^2}{120}
\begin{pmatrix}
(N-2)^2(\mathcal{B}(\stil,\ttil)+\mathcal{B}(\stil,\util) ) \\
\frac{1}{4}(N-4)^2(\mathcal{B}(\stil,\ttil)+\mathcal{B}(\stil,\util) ) \\
4 (\mathcal{B}(\stil,\ttil) +\mathcal{B}(\stil,\util))\\
\mathcal{B}(\stil,\ttil) +\mathcal{B}(\stil,\util) \\
 \hline
\frac{1}{4}(N-2)^2(\mathcal{B}(\stil,\ttil) -\mathcal{B}(\stil,\util))  \\
0  \\
\end{pmatrix}.
\end{equation}
This yields the $s$-channel discontinuity of the one-loop field theory amplitude $V^{(1,0)}_{\bf{a}}$.

According to the discussion in previous sections, higher-derivative terms are instead given by \textit{finite} sums of partial waves, which in turn implies that the associated leading-log coefficients are polynomials. The  first two string corrections are particularly simple. For the leading-log coefficient at order $\ap^6$, we find
\begin{equation}
F^2\vert F^4:\qquad f_{\bf{a}}^{(1,2)} =\zeta_2  \frac{\pi^2}{180}\tilde{s}^3
\begin{pmatrix}
-(N-2)(2N-1) \\
-\frac{1}{2}(N-1)(N-4)\\
4\\
1 \\
 \hline
0   \\
0  \\
\end{pmatrix}.
\end{equation}
Note that this is the only order at one loop for which the antisymmetric contribution is zero. The reason is simple: the amplitude originates from gluing the field-theory amplitude with a constant string correction, which has vanishing antisymmetric contribution and thus projects out the antisymmetric part of the field-theory amplitude. Next, at order  $\ap^7$ we have
\begin{equation}
F^2\vert\partial^2F^4:\qquad f_{\bf{a}}^{(1,3)} = -\zeta_3  \frac{\pi^2}{180}\tilde{s}^3
\begin{pmatrix}
(N-2)^2 \tilde{s} \\
\frac{1}{4}(N-4)^2 \tilde{s}\\
4 \tilde{s}  \\
 \tilde{s} \\
 \hline
\frac{1}{28} (N-2)^2 (\tilde{t}-\tilde{u})  \\
0  \\
\end{pmatrix}.
\end{equation}

The full non-analytic part of the one-loop amplitude is then obtained by adding the crossing images of the above $s$-channel coefficients. This can be equivalently done in the trace-basis or in the projector-basis. In the latter, this is implemented with the help of the crossing matrices, c.f. \eqref{eq:veneziano_full}. To facilitate comparison with the existing literature \cite{Edison:2021ebi,Eberhardt:2022zay}, we present the complete result in the more familiar trace-basis. The single-trace one-loop amplitude reads
\begin{equation}\label{eq:singletraceonel}
\begin{split}
\frac{1}{\ap^3} V_{1234}^{(1)} &= V_{1234}^{(1,0)} - \frac{\zeta_2 \pi^2}{180} \bigg[ (N-3)( \tilde{s}^3 \log(-\tilde{s})+ \tilde{t}^3 \log(-\tilde{t})) + 2  \tilde{u}^3 \log(-\tilde{u})\bigg] \\
&~~ -\frac{\zeta_3 \pi^2 }{1260}\bigg[\tilde{s}^3\big((4N-22)\tilde{s}+(N-2)\tilde{t}\big)\log(-\tilde{s}) +({\tilde{s}\leftrightarrow \tilde{t}})- 14 \tilde{u}^4 \log(-\tilde{u})\bigg] \\ 
&~~ -\frac{\zeta_2^2 \pi^2 }{2^5 \cdot 1575}\bigg[\tilde{s}^3 \big(2(92N{-}219)\tilde{s}^2+(4N{-}9)\tilde{t}^2 +(15{-}8N) \tilde{s}\,\tilde{t}\big)\log(-\tilde{s}) +({\tilde{s}\leftrightarrow \tilde{t}}) \\
&~~ \qquad\qquad\qquad + 2 \tilde{u}^3(40\tilde{s}^2+ 73\tilde{s}\tilde{t}+40\tilde{t}^2) \log(-\tilde{u}) \bigg] \\
&~~   -\frac{\pi^2 }{2^4 \cdot 315}\bigg[\tilde{s}^3 \Big( (N-3) \tilde{s}(3\tilde{s}-\tilde{t})(4\tilde{s}+\tilde{t})
\zeta_2 \zeta_3 + \tfrac{1}{3}\big( (38N{-}208)\tilde{s}^3{+} 6(2N{-}5)\tilde{s}^2\tilde{t}\\
&~~\qquad\qquad\qquad\quad+3 (N{-}4)\tilde{s}\tilde{t}^2 + (N{-}2)\tilde{t}^3 \big)\zeta_5 \Big) \log(-\tilde{s})  +({\tilde{s}\leftrightarrow \tilde{t}})  \\
&~~\qquad\qquad~~ -  2 \tilde{u}^4 \Big((3\tilde{s}+4\tilde{t})(4\tilde{s}+3\tilde{t})
\zeta_2 \zeta_3 -\tilde{s} (22\tilde{s}^2+43 \tilde{s} \tilde{t}+22 \tilde{t}^2)\zeta_5 \Big)      \log(-\tilde{u})\bigg]  \\
&~~  + \mathcal{O}(\alpha'^7) + (\texttt{analytic})\,,
\end{split}
\end{equation}
where $V_{1234}^{(1,0)}$ is the one-loop field-theory amplitude
\begin{equation}
V_{1234}^{(1,0)} =  \frac{\pi^2}{120} \Big( (N-4)\mathcal{I}(\stil,\ttil) -2 (\mathcal{I}(\stil,\util) +\mathcal{I}(\ttil,\util) ) \Big)  \,.
\end{equation}
Here $\mathcal{I}(\stil,\ttil)$ denotes the one-loop box diagram in 10d, with $\text{Disc}_s \mathcal{I}(\stil,\ttil) = \mathcal{B}(\stil,\ttil)$.

For the double-trace terms, we have
\begin{equation}
\begin{split}\label{eq:doubleoneloop}
\frac{1}{\ap^3} V_{12;34}^{(1)} &= V_{12;34}^{(1,0)} - \frac{\zeta_2 \pi^2 }{180}\bigg[2 \tilde{s}^3 \log(-\tilde{s})- \tilde{t}^3 \log(-\tilde{t}) - \tilde{u}^3 \log(-\tilde{u})\bigg]\\
&~~ -\frac{\zeta_3 \pi^2 }{180}\bigg[\tilde{s}^4 \log(-\tilde{s})+\tilde{t}^4\log(-\tilde{t}) + \tilde{u}^4 \log(-\tilde{u}) \bigg] \\
&~~ -\frac{\zeta_2^2 \pi^2 }{2^3 \cdot 1575}\bigg[ \tilde{s}^3(95\tilde{t}^2+95\tilde{u}^2+ 188\tilde{t} \tilde{u})\log(-\tilde{s}) {-} \frac{1}{4}\tilde{t}^3(7\tilde{s}^2+40\tilde{t}^2 + 7 \tilde{s}\tilde{t})\log(-\tilde{t}) +(\tilde{t}\leftrightarrow \tilde{u}) \bigg] \\ 
&~~  - \frac{1}{2^4 \cdot 315}\bigg[2  \tilde{s}^4 \big((4\tilde{t}+3\tilde{u})(4\tilde{u}+3\tilde{t})
\zeta_2 \zeta_3 + (22\tilde{t}^2+22\tilde{u}^2+43 \tilde{t} \tilde{u})\zeta_5 \big)  \log(-\tilde{s}) \\
&~~ \qquad\qquad\quad  + \tilde{t}^4 \big((\tilde{s}-3\tilde{t})(\tilde{s}+4\tilde{t})
\zeta_2 \zeta_3  +  (22\tilde{s}^2+22\tilde{u}^2+43 \tilde{s} \tilde{u})\zeta_5 \big)   \log(-\tilde{t}) +(\tilde{t}\leftrightarrow \tilde{u}) \bigg] \\
&~~  + \mathcal{O}(\alpha'^7) +(\texttt{analytic})\,,
\end{split}
\end{equation}
where the field-theory double-trace term reads
\begin{equation}
V_{12;34}^{(1,0)} =  \frac{\pi^2}{120} \Big( \mathcal{I}(\stil,\ttil) +  \mathcal{I}(\stil,\util)+ \mathcal{I}(\ttil,\util)  \Big)  \,,
\end{equation}
and a few more orders are displayed in an \textit{ancillary file} attached to the submission. Note that the double-trace partial amplitude $V_{12;34}^{(1)}$ does not depend on $N$, in agreement with the interpretation that it arises from the non-planar cylinder topology as discussed in Section \ref{subsec:stringtop} (c.f. Figure \ref{figure:onelooptopologies}).
 
We have checked that the above results \eqref{eq:singletraceonel}-\eqref{eq:doubleoneloop} are in perfect agreement with the imaginary part of the planar cylinder contribution (leading order in $N$ of \eqref{eq:singletraceonel}) computed in  \cite{Edison:2021ebi}, as well as with the worldsheet computations of \cite{Eberhardt:2022zay}.

%%%%%%%%%%%%%%%%%%%%%%%%%%%%%%%%%%%%%%%%%%%%%%%%%%%%%%%%%%%%%%%%%%%%%%%%%%%%%%%%%%%%%%%%%%
\subsubsection{Two loops}
By the same token, we can iterate the procedure to obtain the iterated cut for higher loops. Here we will skip the intermediate steps and just present the results, as the logic is the same as at one loop. The first three orders of the two-loop single-trace amplitude read
\begin{equation}
\begin{split}
\frac{1}{\ap^3}  V_{1234}^{(2)} &= V_{1234}^{(2,0)}  - \frac{\zeta_2 \pi^4 }{2^{9} \cdot 675} \bigg[ (N{-}2) (N{-}5)(\tilde{s}^6\log^2(-\tilde{s}){+}\tilde{t}^6\log^2(-\tilde{t}) {+} 12 \tilde{u}^6 \log^2(-\tilde{u}) \bigg]   \\
&~~ +  \frac{ \zeta_3  \pi^4 }{2^{9} \cdot   33075}\bigg[ (\tilde{s}^6 \big((25 N^2 {-}247 N{+}688)\tilde{s}{+}(N{-}2)^2 \tilde{t}\big)\log^2(-\tilde{s}) +(\tilde{s}\leftrightarrow \tilde{t}) {-} 588\tilde{u}^7  \log^2(-\tilde{u}) \bigg]  \\
&~~  + \frac{ \zeta_2^2  \pi^4 }{2^{14} \cdot   165375} \bigg[\tilde{s}^6 \Big( ( 6952 N^2  - 40354 N+49612)\tilde{s}^2 - (44N^2-167N+170)\tilde{s}\,\tilde{t}\\
&~~\qquad\qquad\qquad\qquad~~ + (4N^2-25N+22)\tilde{t}^2 \Big) \log^2(-\tilde{s})+(\tilde{s}\leftrightarrow \tilde{t})  \\
&~~\qquad\qquad\qquad\quad  +84 \tilde{u}^6(34\tilde{s}^2+ 34\tilde{t}^2 +67 \tilde{s}\tilde{t}) \log^2(-\tilde{u})\bigg]\\
&~~+\mathcal{O}(\ap^9) + (\texttt{lower log's}) + (\texttt{analytic})\,.
\end{split}
\end{equation}
Similarly, for the double-trace terms we obtain
\begin{equation}
\begin{split}
\frac{1}{\ap^3}  V_{12;34}^{(2)} &= V_{12;34}^{(2,0)}+   \frac{\zeta_2 \pi^4 }{2^{8}\cdot 675} \bigg[ (3N{-}7) \tilde{s}^6 \log^2(-\tilde{s})- \tilde{t}^6 \log^2(-\tilde{t})-\tilde{u}^6 \log^2(-\tilde{u}) \bigg] \\
&~~  + \frac{ \zeta_3 \pi^4 }{2^{8}\cdot 675}\bigg[ (3N{-}10) \tilde{s}^7 \log^2(-\tilde{s})+2\tilde{t}^7\log^2(-\tilde{t}) + 2 \tilde{u}^7 \log^2(-\tilde{u}) \bigg] \\
&~~  \frac{ \zeta_2^2 \pi^4 }{2^{13} \cdot 165375} \bigg[ \tilde{s}^6( ( 20928 N-40450 )\tilde{s}^2 + (12N-25)(\tilde{t}^2+\tilde{s}\tilde{t})  )\log^2(-\tilde{s}) \\ 
&~~\qquad\qquad\qquad -\tilde{t}^6(\tilde{s}^2+ 34\tilde{t}^2 + \tilde{s}\tilde{t})\log^2(-\tilde{t}) + (\tilde{t}\leftrightarrow \tilde{u}) \bigg] \\
&~~+\mathcal{O}(\ap^9) + (\texttt{lower log's}) + (\texttt{analytic})\,,
\end{split}
\end{equation}
with the other orientations related by crossing. A few more orders (up to $\ap^{10}$) are given in the \textit{ancillary file}.

Note again that the $N$-dependence of the results is consistent with the discussion based on topology considerations.

%%%%%%%%%%%%%%%%%%%%%%%%%%%%%%%%%%%%%%%%%%%%%%%%%%%%%%%%%%%%%%%%%%%%%%%%%%%%%%%%%%%%%%%%%%
\subsubsection{Three loops}
Lastly, let us also give the first three corrections at three-loop order. 
The single-trace term is given by
\begin{equation}
\begin{split}
\frac{1}{\ap^3}  V_{1234}^{(3)} &=   V_{1234}^{(3,0)}  \\
&~~ -  \frac{ \zeta_2 \pi^6 }{2^{13}  \cdot 91125}\bigg[(N^3{-}11 N^2{+}38 N{-}36 )\tilde{s}^9 \log^3(-\tilde{s}) + (\stil\leftrightarrow\ttil) + 40 \tilde{u}^9\log^3(-\tilde{u}) \bigg]  \\
&~~ -\frac{ \zeta_3  \pi^6 }{2^{13} \cdot 31255875}\bigg[ \tilde{s}^9 \big( 4 \tilde{s}( 43 N^3 {-} 601 N^2  {+} 2917 N {-}5146){+} (N{-}2)^3 \tilde{t}
\big)\log^3(-\tilde{s})\\
&~~\qquad\qquad\qquad\qquad + (\tilde{s}\leftrightarrow \tilde{t})  - 13720 \tilde{u}^{10} \log^3(-\tilde{u}) \bigg]  \\
&~~  +\frac{ \zeta_2^2  \pi^6 }{2^{20} \cdot 156279375}\bigg[ \tilde{s}^9 \Big( \tilde{s}^2 (242248 N^3-2289410 N^2+6720020 N -5821752 )\\
&~~\qquad\qquad\qquad\qquad\qquad  - (188 N^3 - 1111 N^2  + 2182 N - 1476)\tilde{s}\tilde{t}\\ &~~\qquad\qquad\qquad\qquad\qquad -(4N^3-41N^2+122N-60) \tilde{t}^2 \Big)  \log^3(-\tilde{s})\\
&~~\qquad\qquad\qquad\qquad\quad~~ + (\tilde{s}\leftrightarrow \tilde{t})+ 280\tilde{u}^9 (2414\tilde{s}^2+4829\tilde{s}\tilde{t}+2414\tilde{t}^2) \log^3(-\tilde{u}) \bigg]\\
&~~  +\mathcal{O}(\ap^{12}) + (\texttt{lower log's}) + (\texttt{analytic})\,,
\end{split}
\end{equation}
and the double-trace terms reads
\begin{equation}
\begin{split}
\frac{1}{\ap^3}  V_{12;34}^{(3)} &= V_{12;34}^{(3,0)} \\
&~~  -  \frac{\zeta_2 \pi^6 }{2^{12}\cdot 91125} \bigg[(N-2)(7N-19) \tilde{s}^9\log^3(-\tilde{s})- 6\tilde{t}^9\log^3(-\tilde{t}) - 6\tilde{u}^9\log^3(-\tilde{u}) \bigg]   \\
&~~  - \frac{ \zeta_3 \pi^6 }{2^{13} \cdot 3^7 \cdot  5^2 }\bigg[ (7N^2-42N+68)\tilde{s}^{10} \log^3(-\tilde{s})+12 \tilde{t}^{10} \log^3(-\tilde{t}) + 12 \tilde{u}^{10} \log^3(-\tilde{u}) \bigg] \\
&~~  - \frac{ \zeta_2^2 \pi^6 }{2^{19} \cdot 156279375 }\bigg[\tilde{s}^{9} \Big( (1696408 N^2- 6869958 N+ 6721172 )\tilde{s}^2 \\
&~~\qquad\qquad\qquad\qquad\qquad +( 28 N^2{-}123 N{+}122  ) (\tilde{t}^2{+}\tilde{s}\tilde{t})\Big)  \log^3({-}\tilde{s}) \\
&~~\qquad\qquad\qquad\qquad\quad + 42 \Big( \tilde{t}^9 (\tilde{s}^2{+}\tilde{s}\tilde{t}{-}2414\tilde{t}^2)  \log^3({-}\tilde{t})  + (\tilde{t}\leftrightarrow \tilde{u})  \Big)     \bigg] \\
&~~  +\mathcal{O}(\ap^{12}) + (\texttt{lower log's}) + (\texttt{analytic}) \,.
\end{split}
\end{equation}
For a few more orders see the \textit{ancillary file}.

%%%%%%%%%%%%%%%%%%%%%%%%%%%%%%%%%%%%%%%%%%%%%%%%%%%%%%%%%%%%%%%%%%%%%%%%%%%%%%%%%%%%%%%%%%
\section{Conclusions}\label{sec:conclusions}
%%%%%%%%%%%%%%%%%%%%%%%%%%%%%%%%%%%%%%%%%%%%%%%%%%%%%%%%%%%%%%%%%%%%%%%%%%%%%%%%%%%%%%%%%%
In this paper we utilized the partial-wave representation of the tree-level amplitude to compute leading $s$-channel non-analytic terms to multi-loop order for both open and closed superstrings. The orthogonality nature of the partial-wave polynomials -- with respect to the two-particle phase space integral -- allows the computation of the associated unitarity cut to be reduced to sum over products of partial waves. One of the main simplifications in our computation is the fact that in the $\alpha^\prime$-expansion of tree-level superstring amplitudes, the partial-wave expansion is truncated at finite spins for all stringy corrections. This is a reflection of the fact that maximal susy forbids the modification of the cubic couplings in the massless sector. Thus, when computing the non-analytic terms in the $\alpha^\prime$-expansion, the partial-wave sums are always truncated beyond the supergravity/super Yang-Mills contributions. 

The leading trajectory partial waves, defined as the highest-spin partial wave at fixed order in the $\alpha'$-expansion, are the simplest. The resulting leading logs can be resummed into a compact form, reflecting the fact that these operators are renormalized by pure supergravity/super Yang-Mills. On the other hand, it would be interesting to study the renormalization of sub-leading Regge trajectories which will however be more involved as stringy corrections of different genera and $\alpha'$ orders will mix and contribute.

We have also observed an intricate interplay between \textit{sub}-leading log's and higher-genus corrections to \textit{analytic} terms in the low-energy expansion. In particular, the transcendental weight of known higher-genus analytic terms appears to be related across different genera in precisely such a way that the sub-leading log coefficients are of uniform transcendentality. While this is manifest for the leading log's, the possible extension to sub-leading log's is a novel observation, for which we provided some non-trivial evidence. Naturally, it would be desirable to find a more rigorous argument or even proof for these observations, as this could teach us valuable information about the transcendental structure of unprotected higher-derivative terms.\\\vspace{-0.2cm}

The results of our work can serve as boundary data for future computations of higher-genus string amplitudes, or for testing universal structures. For example, the higher-genus generalization of tree-level doubly copy KLT relations~\cite{Kawai:1985xq} has been a long standing challenge. Only recently progress has been made at one-loop~\cite{Stieberger:2022lss}. Our results can serve as a playground for any tentative proposals of such relations at higher genus.

Considering extensions to heterotic and bosonic strings will require several generalizations. Firstly, without maximal susy the external states now carry polarization tensors, and the partial-wave expansion involves matrices of polynomials, see for example~\cite{Caron-Huot:2022jli}. It would be interesting to explore the phase-space integral of the product of these matrices. Secondly, the partial-wave coefficients of the tree-level amplitude are no longer truncated at finite spins even for sub-leading terms in the $\alpha^\prime$-expansion. This is the reflection of new cubic interactions in the massless sectors which introduce new massless poles in the four-point amplitude.

Finally, to capture non-analytic terms beyond the leading logs, one will require efficient ways to compute unitarity cuts beyond two-particle thresholds. For example, it would be interesting to work out orthogonal polynomials associated with three-particle phase-space integrals. The solutions to these challenges will have wide applications for general EFTs, and thus new efficient approaches are extremely desirable. 

%%%%%%%%%%%%%%%%%%%%%%%%%%%%%%%%%%%%%%%%%%%%%%%%%%%%%%%%%%%%%%%%%%%%%%%%%%%%%%%%%%%%%%%%%
\section*{Acknowledgments}
%%%%%%%%%%%%%%%%%%%%%%%%%%%%%%%%%%%%%%%%%%%%%%%%%%%%%%%%%%%%%%%%%%%%%%%%%%%%%%%%%%%%%%%%%%
We are very grateful to Oliver Schlotterer for helpful discussions and for providing useful comments at multiple stages of this work. We also thank Sebastian Mizera, Lorenz Eberhardt, and Hikaru Kawai for useful discussions. HP is grateful to Silviu Pufu and Princeton University for kind hospitality during part of this work. HP acknowledges support from the FWO grant G094523N. YtH is supported by the National Science and Technology Council (NSTC) of Taiwan grant 112-2628-M-002 -003 -MY3, and MS is supported by the NSTC through the grant 110-2112-M-002-006-.

%%%%%%%%%%%%%%%%%%%%%%%%%%%%%%%%%%%%%%%%%%%%%%%%%%%%%%%%%%%%%%%%%%%%%%%%%%%%%%%%%%%%%%%%%%
%%%%%%%%%%%%%%%%%%%%%%%%%%%%%%%%%%%%%%%%%%%%%%%%%%%%%%%%%%%%%%%%%%%%%%%%%%%%%%%%%%%%%%%%%%
\appendix
%%%%%%%%%%%%%%%%%%%%%%%%%%%%%%%%%%%%%%%%%%%%%%%%%%%%%%%%%%%%%%%%%%%%%%%%%%%%%%%%%%%%%%%%%%
\section{Partial-wave coefficients of higher-genus analytical terms}\label{app:pw_coeffs}
%%%%%%%%%%%%%%%%%%%%%%%%%%%%%%%%%%%%%%%%%%%%%%%%%%%%%%%%%%%%%%%%%%%%%%%%%%%%%%%%%%%%%%%%%%
Here we list the partial-wave coefficients extracted from the higher-genus analytical contributions listed in Section \ref{subsec:known_analytic_terms}. For the genus-1 terms in equation \eqref{eq:higher1}
\begin{align}
\begin{split}
	\R^4_1:\qquad \epsilon_\ell &= \frac{\tilde{s}^3}{2^{11}\cdot315\pi^2}\,\delta_{\ell,0}\,,\\
	\partial^6\R^4_1:\qquad \epsilon_\ell &= \frac{\tilde{s}^6\zeta_3}{2^{16}\cdot2835 \pi^2}\,\Big(\delta_{\ell,0}-\frac{1}{44}\delta_{\ell,2}\Big)\,,\\
	\partial^{10}\R^4_1:\qquad \epsilon_\ell &= \frac{493\stil^8\zeta_5}{2^{22}\cdot467775\pi^2}\,\Big(\delta_{\ell,0}-\frac{4}{221}\delta_{\ell,2}-\frac{1}{2210}\delta_{\ell,4}\Big)\,.\qquad~~
\end{split}
\end{align}
For the genus-2 terms from equation \eqref{eq:higher2}
\begin{align}
\begin{split}
	\partial^4\R^4_2:\qquad \epsilon_\ell &= \frac{7\tilde{s}^5}{2^{14}\cdot127575}\,\Big(\delta_{\ell,0}+\frac{1}{154}\delta_{\ell,2}\Big)\,,\\
	\partial^6\R^4_2:\qquad \epsilon_\ell &= \frac{\tilde{s}^6}{2^{16}\cdot14175}\,\Big(\delta_{\ell,0}-\frac{1}{44}\delta_{\ell,2}\Big)\,,\\
	\partial^{12}\R^4_2:\qquad \epsilon_\ell &= \frac{\tilde{s}^9}{2^{12}\cdot45045}\Big((1624D_1+195D_2)\,\delta_{\ell,0}\\
		&\qquad\qquad\qquad~~+\frac{68D_1-15D_2}{2}\,\delta_{\ell,2}\\
		&\qquad\qquad\qquad~~+\frac{112D_1+51D_2}{170}\,\delta_{\ell,4}+\frac{2}{323}D_1\,\delta_{\ell,6}\Big)\,,
\end{split}
\end{align}
and lastly for the genus-3 term given in equation \eqref{eq:higher3}
\begin{align}
	\partial^6\R^4_3:\qquad \epsilon_\ell &= \frac{\tilde{s}^6\pi^2}{2^{15}\cdot8037225}\,\Big(\delta_{\ell,0}-\frac{1}{44}\delta_{\ell,2}\Big)\,.\qquad\qquad\qquad\qquad
\end{align}

%%%%%%%%%%%%%%%%%%%%%%%%%%%%%%%%%%%%%%%%%%%%%%%%%%%%%%%%%%%%%%%%%%%%%%%%%%%%%%%%%%%%%%%%%%
\section{Crossing matrices for $SO(N)$}\label{sec:crossingmatrices}
%%%%%%%%%%%%%%%%%%%%%%%%%%%%%%%%%%%%%%%%%%%%%%%%%%%%%%%%%%%%%%%%%%%%%%%%%%%%%%%%%%%%%%%%%%
In this appendix we record the expressions for the $SO(N)$ crossing matrices, which are needed for the  implementation of crossing transformations in the projector basis.
The $s$-channel crossing matrix $F_s$ (corresponding to $1\leftrightarrow2$ exchange) is diagonal and measures the parity of the corresponding irrep:
\begin{equation}
F_s = \texttt{diag} \{ 1,1,1,1,-1,-1  \} \,.
\end{equation}
On the other hand, obtaining the $t$- and $u$-channel crossing matrices $F_t$ and $F_u$ (corresponding to $1\leftrightarrow4$ and $1\leftrightarrow3$ exchange) requires a computation. They are computed via (see e.g. \cite{cvitanovic2008group,Isaev:2020kwc} for useful tools)
\begin{equation}
 (F_t)_{{\bf{a}}{\bf{b}}} = \frac{1}{\text{dim}({\bf{a}})}	\mathbb{P}_\mathbf{a}^{I_1I_2I_3I_4}\mathbb{P}_\mathbf{b}^{I_4I_2I_3I_1}\,,\qquad  (F_u)_{{\bf{a}}{\bf{b}}} = \frac{1}{\text{dim}({\bf{a}})}	\mathbb{P}_\mathbf{a}^{I_1I_2I_3I_4}\mathbb{P}_\mathbf{b}^{I_3I_2I_1I_4}\,,
\end{equation}
from which one obtains:
\begin{equation}
F_t=\begin{pmatrix}
\frac{2}{N(N - 1)} & \frac{N + 2}{N} & \frac{(N {-} 3) (N {-} 2) }{12} & \frac{(N - 3)(N + 1)(N + 2)}{6(N - 1)} & -1 & -\frac{(N {-} 3)(N {+} 2) }{4}\\
\frac{2}{N(N - 1)} & \frac{N^2 - 8}{2N(N - 2)} & \frac{3 - N}{6} & \frac{(N - 4)(N - 3)(N + 1)}{6(N - 2)(N - 1)} & -\frac{N - 4}{2(N - 2)} & \frac{N - 3}{N - 2} \\
\frac{2}{N(N - 1)} & -\frac{2(N + 2)}{N(N - 2)} & \frac{1}{6} & \frac{(N + 1)(N + 2)}{3(N - 2)(N - 1)} & -\frac{2}{N - 2} & \frac{N + 2}{2(N - 2)} \\
\frac{2}{N(N - 1)} & \frac{N - 4}{N(N - 2)} & \frac{1}{6} & \frac{N^2 - 6N + 11}{3(N - 2)(N - 1)} & \frac{1}{N - 2} & \frac{N - 4}{2(N - 2)} \\
-\frac{2}{N(N - 1)} & -\frac{(N - 4)(N + 2)}{2N(N - 2)} & \frac{3 - N}{6} & \frac{(N - 3)(N + 1)(N + 2)}{6(N - 2)(N - 1)} & \frac{1}{2} & 0 \\
-\frac{2}{N(N - 1)} & \frac{4}{N(N - 2)} & \frac{1}{6} & \frac{(N - 4)(N + 1)}{3(N - 2)(N - 1)} & 0 & \frac{1}{2}
\end{pmatrix},
\end{equation}
\begin{equation}
F_u=\begin{pmatrix}
\frac{2}{N(N - 1)} & \frac{N + 2}{N} & \frac{(N {-} 3) (N {-} 2) }{12} & \frac{(N - 3)(N + 1)(N + 2)}{6(N - 1)} & 1 & \frac{(N {-} 3)(N {+} 2)}{4} \\
\frac{2}{N(N - 1)} & \frac{N^2 - 8}{2N(N - 2)} & \frac{3 - N}{6} & \frac{(N - 4)(N - 3)(N + 1)}{6(N - 2)(N - 1)} & \frac{N - 4}{2(N - 2)} & -\frac{N - 3}{N - 2} \\
\frac{2}{N(N - 1)}& -\frac{2(N + 2)}{N(N - 2)} & \frac{1}{6} & \frac{(N + 1)(N + 2)}{3(N - 2)(N - 1)} & \frac{2}{N - 2} & -\frac{N + 2}{2(N - 2)} \\
\frac{2}{N(N - 1)} & \frac{N - 4}{N(N - 2)} & \frac{1}{6} & \frac{N^2 - 6N + 11}{3(N - 2)(N - 1)} & -\frac{1}{N - 2} & -\frac{N - 4}{2(N - 2)} \\
\frac{2}{N(N - 1)} & \frac{(N - 4)(N + 2)}{2N(N - 2)} & \frac{N-3}{6}& -\frac{(N - 3)(N + 1)(N + 2)}{6(N - 2)(N - 1)} & \frac{1}{2} & 0 \\
\frac{2}{N(N - 1)}& -\frac{4}{N(N - 2)} & -\frac{1}{6} & -\frac{(N - 4)(N + 1)}{3(N - 2)(N - 1)} & 0 & \frac{1}{2}
\end{pmatrix}.
\end{equation}
As a consistency check, one can verify that the above matrices satisfy the relations $F_s^2=F_t^2=F_u^2= 1$, and $F_t  F_u  F_t=F_u  F_t  F_u=F_s$, which are implied by crossing symmetry.

%%%%%%%%%%%%%%%%%%%%%%%%%%%%%%%%%%%%%%%%%%%%%%%%%%%%%%%%%%%%%%%%%%%%%%%%%%%%%%%%%%%%%%%%%%
%%%%%%%%%%%%%%%%%%%%%%%%%%%%%%%%%%%%%%%%%%%%%%%%%%%%%%%%%%%%%%%%%%%%%%%%%%%%%%%%%%%%%%%%%%
\bibliography{references}
\bibliographystyle{JHEP}
\end{document}